%% file: main.tex
\documentclass[letterpaper,twocolumn,10pt]{article}
\usepackage{usenix2019_v3}

\usepackage{tikz}

\usepackage{cite}
\usepackage{amsmath,amssymb,amsfonts}
\usepackage{graphicx}
\usepackage{textcomp}
\usepackage{xcolor}
\usepackage{subcaption}
\usepackage{multirow}
\usepackage{pifont}
\usepackage{booktabs}
\usepackage{siunitx}

\usepackage{soul}

\usepackage{breakurl}

\newlength{\bibitemsep}
\newlength{\bibparskip}
\let\oldthebibliography\thebibliography
\renewcommand\thebibliography[1]{%
  \oldthebibliography{#1}%
  \setlength{\parskip}{\bibitemsep}%
  \setlength{\itemsep}{\bibparskip}%
}

\input{math_commands.tex}

\def\BibTeX{{\rm B\kern-.05em{\sc i\kern-.025em b}\kern-.08em
    T\kern-.1667em\lower.7ex\hbox{E}\kern-.125emX}}

\newcommand{\etal}{\textit{et al.}}
\newcommand{\ie}{\textit{i}.\textit{e}.}
\newcommand{\eg}{\textit{e}.\textit{g}.}

\usepackage{algorithm, algorithmicx}
\usepackage{algpseudocode}

\let\oldReturn\Return
\renewcommand{\Return}{\State\oldReturn}
\algnewcommand\algorithmicforeach{\textbf{for each}}
\algdef{S}[FOR]{ForEach}[1]{\algorithmicforeach\ #1\ \algorithmicdo}

\newcommand{\attacktrials}{\text{A\textsc{ttack}}\text{T\textsc{rials}}}

\iffalse

\newcommand{\ncaption}[1]{\caption{#1}}
\newcommand{\nsection}[1]{\section{#1}}
\newcommand{\nsubsection}[1]{\subsection{#1}}

\else

\newcommand{\ncaption}[1]{\caption{#1}}
\newcommand{\nsection}[1]{\vspace{-0.25cm}\section{#1}\vspace{-0.15cm}}
\newcommand{\nsubsection}[1]{\vspace{-0.4cm}\subsection{#1}\vspace{-0.1cm}}

\fi

\usepackage{enumitem}

\begin{document}

\date{}

\title{\Large \bf Drift with Devil: Security of Multi-Sensor Fusion based Localization in High-Level Autonomous Driving under GPS Spoofing \\
\large (Extended Version)}

\author{
{\rm Junjie Shen}\\
UC Irvine\\
junjies1@uci.edu
\and
{\rm Jun Yeon Won}\\
UC Irvine\\
junyeonw@uci.edu
\and
{\rm Zeyuan Chen}\\
UC Irvine\\
zeyuac4@uci.edu
\and
{\rm Qi Alfred Chen}\\
UC Irvine\\
alfchen@uci.edu
}

\maketitle

\input{abstract}

\input{intro}
\input{background}

\input{threat_model}
\input{security_analysis}

\input{attack_design}

\input{evaluation}
\input{practical_attack}
\input{offline_profile}
\input{discussion}

\input{related_work}

\input{conclusion}
\input{acknowledgement}

\bibliographystyle{ieeetr}
{
\footnotesize
\bibliography{reference,msf_survey}
}

\appendix

\include{appendix}

\end{document}

%% file: math_commands.tex
\usepackage{amsmath,amsfonts,bm}

\def\eqref#1{equation~\ref{#1}}

\def\1{\bm{1}}

\DeclareMathAlphabet{\mathsfit}{\encodingdefault}{\sfdefault}{m}{sl}
\SetMathAlphabet{\mathsfit}{bold}{\encodingdefault}{\sfdefault}{bx}{n}

%% file: abstract.tex
\begin{abstract}
For high-level Autonomous Vehicles (AV), localization is highly security and safety critical. One direct threat to it is GPS spoofing, but fortunately, AV systems today predominantly use Multi-Sensor Fusion (MSF) algorithms that are generally believed to have the potential to practically defeat GPS spoofing. However, no prior work has studied whether today's MSF algorithms are indeed sufficiently secure under GPS spoofing, especially in AV settings. In this work, we perform the first study to fill this critical gap. As the first study, we focus on a production-grade MSF with both design and implementation level representativeness, and identify two AV-specific attack goals, off-road and wrong-way attacks.

To systematically understand the security property, we first analyze the upper-bound attack effectiveness, and discover a take-over effect that can fundamentally defeat the MSF design principle. We perform a cause analysis and find that such vulnerability only appears dynamically and non-deterministically. Leveraging this insight, we design FusionRipper, a novel and general attack that opportunistically captures and exploits take-over vulnerabilities. We evaluate it on 6 real-world sensor traces, and find that FusionRipper can achieve at least 97\% and 91.3\% success rates in all traces for off-road and wrong-way attacks respectively. We also find that it is highly robust to practical factors such as spoofing inaccuracies. To improve the practicality, we further design an offline method that can effectively identify attack parameters with over 80\% average success rates for both attack goals, with the cost of at most half a day. We also discuss promising defense directions.

\end{abstract}

%% file: intro.tex
\vspace{-0.1in}
\nsection{Introduction}\label{sec:intro}
\vspace{-0.07in}

Today, various companies are developing high-level self-driving cars~\cite{av-companies} such as Level-4 Autonomous Vehicles (AV)~\cite{sae2018}, and some of them are already providing services on public roads such as self-driving taxi from Google's Waymo One~\cite{waymo-one} and self-driving trucks from TuSimple~\cite{tusimple-truck}. To enable such high-level driving automation, the Autonomous Driving (AD) system in an AV needs to not only perform the perception of surrounding obstacles, but also \textit{centimeter-level localization} of its own global positions on the map~\cite{levinson2007map, reid2019localization}. Such localization function is highly security and safety critical in the AV context, since positioning errors can directly cause an AV to drive off road or onto a wrong way. Since in high-level AD systems the perception module is only designed for obstacle detection and the localization module is in full charge of identifying road deviations~\cite{udacity_av_nd, udacity_av_apollo, coursera_av, apollo, autoware}, even when the perception module is functioning perfectly, it cannot prevent a variety of road hazards specific to localization errors such as driving off road to hit road curbs, falling down the highway cliff, or being hit by other vehicles that fail to yield, especially when the AV is on the wrong way. 
However, recent security research in AD systems concentrates on AD perception, \eg, malicious stickers on traffic signs~\cite{eykholt2018robust, eykholt2018physical, zhao2019seeing, cao2019adversarial}, which leaves the security of AD localization an open problem.

For outdoor localization in general, GPS is the \textit{de facto} location source, and thus a direct threat to it is GPS spoofing, a long-existing but still unsolved security problem with practicality proven on a wide range of end systems~\cite{popperccs11,utaustinspoofer, zeng2018all,narain2018security,franceschi2012drone,kerns2014unmanned,spoof_tesla,spoofyacht, noh2019tractor}, including low-autonomy AVs such as Tesla cars~\cite{spoof_tesla}. Fortunately, to achieve robust localization, real-world high-level AD systems today predominantly use Multi-Sensor Fusion (MSF) algorithms that combine GPS input with position inputs from other sensors, typically IMU (Inertial Measurement Unit) and LiDAR (Light Detection and Ranging)~\cite{udacity_av_nd,wan2018robust,gao2015ins,suhr2016sensor,tao2013mapping,schreiber2016vehicle,de2017survey,soloviev2008tight,lee2015gps,kelly2011visual}. Since in such design GPS input alone can not dictate the localization output, it is generally believed to have the potential to practically defeat GPS spoofing~\cite{spoofyacht,zeng2018all,lee2017attack,davidson2016controlling,albrektsen2018robust,jafarnia2012gps}. However, state-of-the-art MSF algorithms are mainly designed for improving accuracy and robustness, instead of security. This thus makes it largely unclear how secure they can be under GPS spoofing.
Given its widespread use in AVs and high importance to road safety, it is thus imperative to systematically understand this as early as possible.

To fill this critical research gap, in this work we perform the first study on the security property of MSF-based localization in AV settings. As the very first study in this direction, we focus on GPS spoofing as the attack vector since it is one of the most mature attack vectors to the MSF input sources. We focus on a production-grade MSF implementation, Baidu Apollo MSF (BA-MSF), due to its high representativeness in both design (KF-based MSF) and implementation (centimeter-level accuracy evaluated by real-world AV fleet), which will be detailed later in \S\ref{sec:msf}.
We consider the attack goal as using GPS spoofing to cause large \textit{lateral} deviations in the MSF output, \ie, deviating to the left or right. This can cause the AV to drive off road or onto a wrong way, which we call \textit{off-road attack} and \textit{wrong-way attack} respectively.

To systematically understand the security property, we first analyze the upper-bound attack effectiveness via a dynamic blackbox analysis since BA-MSF is released in the binary form. We find that in the real-world trace, the majority (71\%) of even such upper-bound attack results can only cause less than 50 cm deviation, which is far from causing either off-road or wrong-way attacks (need over 90 cm and 2.4 m respectively). This shows that MSF can indeed generally enhance the security against GPS spoofing. Interestingly, we also observe that there still exist a few upper-bound attack results that can cause over 2 meters deviations. For all of them, we find that GPS spoofing is able to cause \textit{exponential growths} of deviations. This allows the spoofed GPS to become the dominating input source in the fusion process and eventually cause the MSF to reject other input sources, which thus \textit{fundamentally defeats the design principle of MSF}. In this paper, we call it a \textit{take-over effect}. We then perform a cause analysis and find that this only appears when the MSF is in relatively \textit{unconfident} periods due to a combination of dynamic and non-deterministic real-world factors such as sensor noises and algorithm inaccuracies.

Such take-over vulnerabilities are highly attractive for attackers since they can exploit the exponential deviation growths to achieve \textit{arbitrary} deviation goals. However, as discovered earlier, the vulnerable periods are created dynamically and non-deterministically.
Thus, we design \textit{FusionRipper}, a novel and general attack that opportunistically captures and exploits take-over vulnerabilities with 2 stages: (1) \textit{vulnerability profiling}, which measures when vulnerable periods appear, and (2) \textit{aggressive spoofing}, which performs exponential spoofing to exploit the take-over opportunity.

We implement FusionRipper and evaluate it on 6 real-world sensor traces from Apollo and the KAIST Complex Urban dataset.
Our results show that when the attack can last 2 minutes, there \textit{always} exists a set of attack parameters for FusionRipper to achieve \textit{at least} 97\% and 91.3\% success rates in \textit{all} traces for the off-road and wrong-way attacks respectively, with less than 35 seconds success time on average. To understand the attack practicality, we evaluate it with practical factors such as (1) spoofing inaccuracies, and (2) AD control taking effect, and find that for both cases the attack success rates are affected by less than 4\%. Attack demos showing the end-to-end attack impact are available at
\textbf{\url{https://sites.google.com/view/cav-sec/fusionripper}}.

In addition, we observe that the attack effectiveness is sensitive to the selection of the attack parameters. Thus, to improve the practicality, we further design an offline attack parameter profiling method that can collect effective parameters without causing obvious safety problems during such profiling to stay stealthy. Our results on real-world traces show that our method can effectively identify attack parameters with 84.2\% and 80.7\% success rates for off-road and wrong-way attacks respectively, with the profiling cost of at most half a day.

Considering the critical role of localization for safe and correct AV driving, the discovered attack against the state-of-the-art MSF algorithm requires immediate attention and defense discussion. To facilitate this, we also discuss both long-term and short-term defense directions.

In summary, this work makes the following contributions:

\vspace{-\topsep}
\begin{itemize}
\setlength{\itemsep}{0pt}
\setlength{\parskip}{0pt}
\item We perform the first security study on MSF-based localization in high-level AV settings under GPS spoofing. We focus on a production-grade MSF with both design and implementation level representativeness, and identify two attack goals specific to the AV settings.

\item We analyze the upper-bound attack effectiveness, and discover a take-over effect that can fundamentally defeat the MSF design principle. We further perform a cause analysis and find that such vulnerability only appears dynamically and non-deterministically.

\item We design FusionRipper, a novel and general attack that opportunistically captures and exploits the take-over vulnerability we discover. We evaluate it on 6 real-world sensor traces, and find that it can achieve high effectiveness (over 97\% and 91.3\% success rates) for both off-road and wrong-way attacks. We also find that such high effectiveness is robust to various practical factors.

\item To improve the attack practicality, we further design an offline attack parameter profiling method that can effectively identify attack parameters with 84.2\% and 80.7\% success rates for off-road and wrong-way attacks respectively, with the profiling cost of at most half a day. We also discuss promising defenses directions.

\vspace{-\topsep}
\end{itemize}

%% file: background.tex
\vspace{-0.1in}
\nsection{Background}\label{sec:background}
\vspace{0.08in}

\nsubsection{AD Localization and Multi-Sensor Fusion}\label{sec:msf}
\vspace{-0.05in}

\begin{figure}[tbp]
\centering
\includegraphics[width=.95\columnwidth]{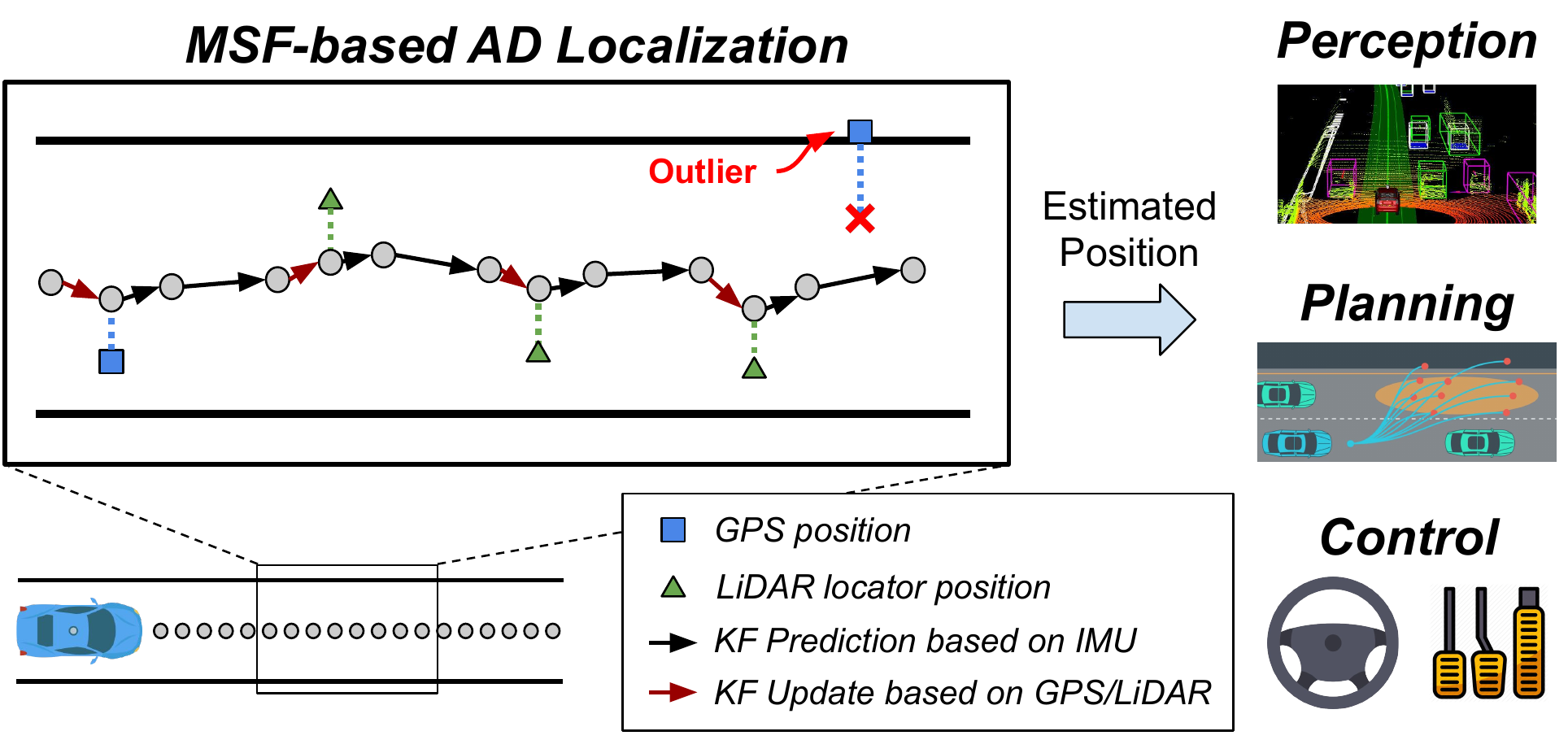}
\vspace{-0.15in}
\ncaption{MSF-based localization and its use in high-level AD systems.}
\label{fig:av_msf}
\vspace{-0.25in}
\end{figure}

In real-world high-level (\eg, Level 4~\cite{sae2018}) AD system design, localization is a critical module that needs to compute global vehicle positions on the map in the real time based on positioning sensor inputs~\cite{udacity_av_nd, udacity_av_apollo, coursera_av, apollo, autoware}. As shown in Fig.~\ref{fig:av_msf}, its output is used by various other modules in the AD system, \eg, the perception module for detecting obstacles, the planning module for driving decision making, and the control module for executing these decisions. Such direct impact on various critical decision making steps in AV driving thus makes localization outputs highly security and safety critical.

To ensure safe and correct driving, AD localization needs to not only have \textit{centimeter-level} accuracy to localize the AV at traffic lane level~\cite{reid2019localization, ega_requirement_report, levinson2007map}, but also have high robustness under various road and weather conditions~\cite{ega_requirement_report}. Thus, Multi-Sensor Fusion (MSF) based design has become the mainstream in both academia and industry since it can fuse results from multiple independent positioning sensors, typically GPS, IMU, and LiDAR, and thus produce results with overall higher accuracy and robustness~\cite{udacity_av_nd,udacity_av_apollo,coursera_av,wan2018robust,gao2015ins,suhr2016sensor,tao2013mapping,schreiber2016vehicle,de2017survey,soloviev2008tight,lee2015gps,kelly2011visual}. For example, modern AV-grade GPS receivers can achieve centimeter-level positioning accuracy with the error correction from ground stations~\cite{gpsdatasheet}. However, GPS signal quality can be easily degraded due to natural phenomena such as atmosphere delays and multi-path effect~\cite{hofmann2007gnss}.
LiDAR-based localization algorithms, or LiDAR locators~\cite{levinson2010robust,gao2015ins, icp,ndt}, match laser scans to pre-generated ones in a High Definition Map (HD Map)~\cite{hdmaps} in order to provide highly accurate positioning.
However, the performance of such matching is susceptible to poor weather conditions such as rain or fog and the outdatedness of the HD Map.
Thus, the goal of MSF is to leverage the strengths of these different sources while compensating their weaknesses.

\textbf{Kalman Filter (KF) based MSF and its representativeness.} Among MSF-based localization algorithms for AD systems, KF-based MSF is adopted most extensively in both academia and industry~\cite{wan2018robust,gao2015ins,tao2013mapping,schreiber2016vehicle,soloviev2008tight,lee2015gps,kelly2011visual}, and shown to have the state-of-the-art performance~\cite{wan2018robust}.
To concretely show its representativeness, we survey the MSF-based localization papers from top-tier robotics conferences~\cite{csrankings} in the most recent 2 years (2018, 2019). As shown in Table~\ref{tbl:msf_survey}, 14 (77.8\%) of the total 18 papers adopt KF-based MSF, showing a clear predominance in today's MSF designs. Such representativeness can also be shown by the fact that it is taught in all Self-Driving Car courses from Udacity~\cite{udacity_av_nd, udacity_av_apollo} and Coursera~\cite{coursera_av}.

KF is a Bayesian filter that calculates an \textit{optimal} state distribution with the \textit{lowest} uncertainty from the sensor measurement distributions.
In the context of AD localization, the state is composed of the vehicle's \textit{position}, \textit{velocity}, and \textit{attitude} (PVA) and their \textit{uncertainties} (or co-variance or variance matrices).
Specifically, KF iteratively applies two steps: \textit{prediction} (Eq.~\ref{eq:kf_pred}) and \textit{update} (Eq.~\ref{eq:kf_update}).
In the $k$-th iteration, the inputs are the previous iteration's KF state $\mathbf{\hat{x}}_{k-1}$ and its state co-variance matrix $\mathbf{\hat{P}}_{k-1}$, which describes the state uncertainty. In the prediction, the acceleration and angular velocity from IMU are integrated in $\mathbf{F}_{k}$ to generate $\mathbf x_k$ and $\mathbf P_k$, which are an intermediate KF state and its co-variance. Next, the update step takes the measurement $\mathbf z_k$ and its uncertainty $\mathbf{R}_k$ from GPS or LiDAR locator, and first use $\mathbf{R}_k$ to calculate Kalman gain $\mathbf{K}_k$. $\mathbf{K}_k$ is then used as a weight to determine how much of the difference between $\mathbf{z}_k$ and $\mathbf{x}_k$ is updated to the new state $\mathbf{\hat{x}}_{k}$, and how much of $\mathbf P_k$ is updated to the new state co-variance $\mathbf{\hat{P}}_k$. In the equations, $\mathbf Q$ and $\mathbf H$ are typically constant matrices, with the former used for tuning the system and the latter for mapping the state space to the measurement space.
\vspace{-0.05in}
\begin{equation}
\small
\begin{aligned}
    \mathbf{x}_{k} &= \mathbf{F}_{k} \mathbf{\hat{x}}_{k-1} \\
    \mathbf{P}_k &= \mathbf{F}_{k} \mathbf{\hat{P}}_{k-1} \mathbf{F}_{k}^T + \mathbf{Q} \\
\end{aligned}
\label{eq:kf_pred}
\end{equation}

\begin{equation}
\small
\begin{aligned}
    \mathbf{\hat{x}}_k &= \mathbf{x}_k + \mathbf{K}_k (\mathbf{z}_k - \mathbf{H} \mathbf{x}_k)\\
    \mathbf{\hat{P}}_k &= \mathbf{P}_k - \mathbf{K}_k \mathbf{H} \mathbf{P}_k\\
    \mathbf{K}_k &= \mathbf{P}_k \mathbf{H}^T (\mathbf{H} \mathbf{P}_k \mathbf{H}^T + \mathbf{R}_k)^{-1} \\
\end{aligned}
\label{eq:kf_update}
\end{equation}

Fig.~\ref{fig:av_msf} shows an example of the KF operations. In the prediction step, the acceleration and angular velocity from IMU are integrated in the KF to generate an intermediate state (black arrows in Fig.~\ref{fig:av_msf}). In the update step, KF takes the position measurements from GPS or LiDAR locator, and updates a fraction of it to the KF state based on the uncertainties of the KF state and the measurement. A larger KF state uncertainty or a smaller measurement uncertainty will cause more updates to the KF state.

\begin{table}[tbp]
\footnotesize
\centering
\ncaption{Survey of MSF-based localization designs in papers published in top-tier robotics conferences (IROS, ICRA, and RSS)~\cite{csrankings} in the most recent 2 years (2018 and 2019).}
\label{tbl:msf_survey}
\vspace{-0.1in}
\setlength{\tabcolsep}{2.7pt}
\begin{tabular}{@{}ccccc@{}}
\toprule
\multicolumn{2}{c}{MSF Design} & \multirow{2}{*}{Papers} & \multicolumn{2}{c}{\multirow{2}{*}{Percentage}} \\ \cmidrule(r){1-2}
Category & Name &  & \multicolumn{2}{c}{} \\ \midrule
\multirow{3}{*}{KF-based} & Linear KF & \cite{piperakis2019outlier,zuo2019lic,zuo2019visual,miiller2018robust,eckenhoff2019multi,arana2019efficient,wan2018robust} & 7/18 (38.9\%) & \multirow{3}{*}{14/18 (77.8\%)} \\
 & Extended KF & \cite{allak2019covariance,brossard2019learning,gosala2019redundant,zhang2018pirvs} & 4/18 (22.2\%) &  \\
 & Unscented KF & \cite{brossard2018unscented,poggenhans2018precise,arnold2018robust} & 3/18 (16.7\%) &  \\ \midrule
\multirow{3}{*}{Others} & Particle Filter & \cite{zhang2019localization} & 1/18 (5.6\%) & \multirow{3}{*}{4/18 (22.2\%)} \\
 & Graph Optimization & \cite{mascaro2018gomsf,geneva2018asynchronous} & 2/18 (11.1\%) &  \\
 & Neural Network & \cite{chame2018reliable} & 1/18 (5.6\%) &  \\ \bottomrule
\end{tabular}
\vspace{-0.25in}
\end{table}

\textbf{Outlier detection.} To prevent KF state from being easily disrupted by occasional measurements that are too noisy in the real world, the KF update is usually bounded by an \textit{outlier detector}. Fig.~\ref{fig:av_msf} shows an example where a GPS measurement is discarded since its position deviates too much from the KF state.
Chi-squared test (Eq.~\ref{eq:nis}) is one of the most widely used outlier detectors for KF~\cite{schreiber2016vehicle,kelly2011visual,piche2016online}, which considers a measurement $\mathbf{z}_k$ as an outlier if the Chi-squared test value $\chi^2_k$ is larger than a statistical significance threshold (usually 3.841~\cite{croarkin2006nist}). An outlier measurement can be either discarded or partially updated.
\vspace{-0.05in}
\begin{equation}
\small
\begin{aligned}
\chi^2_{k} &= (\mathbf{z}_k - \mathbf{H} \mathbf{x}_{k})^{T} \mathbf{S}_{k}^{-1} (\mathbf{z}_k - \mathbf{H} \mathbf{x}_{k}) \\
\mathbf{S}_{k} &= \mathbf{H} \mathbf{P}_{k} \mathbf{H}^T + \mathbf{R}_{k} \\
\end{aligned}
\label{eq:nis}
\end{equation}

\textbf{Targeted MSF implementations and representativeness.} In this paper, we perform our security study on concrete MSF implementations for practicality and realism. In particular, our main target is an MSF design and implementation from the Baidu Apollo team, which we call \textit{BA-MSF}. It is published in ICRA 2018~\cite{wan2018robust}, a top-tier robotics conference~\cite{csrankings}, and follows the KF-based MSF design using high-end GPS, LiDAR, and IMU, with the Chi-squared test as the outlier detector conforming to the common practice~\cite{schreiber2016vehicle,kelly2011visual,piche2016online}. As described earlier, such design is the most representative in today's MSF-based AD localization (Table~\ref{tbl:msf_survey}).

Besides its design, the implementation of BA-MSF is also highly representative in today's MSF-based AD localization: it has been tested using a large AV fleet in various challenging scenarios such as urban downtown, highways, and tunnels~\cite{wan2018robust}, and shown the \textit{highest} localization accuracy (0.054 m) among \textit{all} MSF-based localization papers (including both KF-based and non KF-based) in the top-tier robotics conferences~\cite{csrankings} of the most recent 2 years. Today, it is already adopted in Baidu Apollo~\cite{apollo}, a production-grade AD system currently providing self-driving taxi services in China~\cite{baidu-apollo-go}.

Besides BA-MSF, we also consider two other publicly-available KF-based MSFs for generality evaluations (\S\ref{sec:generality}). We follow the common parameter tuning process~\cite{groves2015principles} but can only reach at most 1-2 meter accuracy, which is far from the centimeter-level accuracy required by AD systems~\cite{levinson2007map, reid2019localization}. Thus, in the majority of our experiments, we target BA-MSF as it is much more representative for AD systems.

\vspace{-0.05in}
\nsubsection{GPS Spoofing and the Practicality} \label{sec:spoofing}
\vspace{-0.05in}

GPS spoofing has been a fundamental problem for civilian GPS systems due to the lack of signal authentication in the infrastructure. In GPS spoofing, the attacker transmits fabricated GPS signals with stronger power than the authentic ones, and thus causes the victim receiver to lock onto the attacker's signals and resolve positions controlled by the attacker. GPS spoofing has been proven feasible theoretically~\cite{popperccs11} and empirically~\cite{utaustinspoofer}. So far, it has been
demonstrated on various end systems such as smartphones~\cite{zeng2018all,narain2018security}, drones~\cite{franceschi2012drone,kerns2014unmanned}, yachts~\cite{spoofyacht}, and recently also low-level AVs such as Tesla cars~\cite{spoof_tesla}.
Recently, a year-long investigation identified 9,883 spoofing events that affected 1,311 civilian vessel systems in Russia since 2016~\cite{utaustin_russia_report}. Although GPS spoofers are illegal to be sold in the U.S., they can be made cheaply from commercial off-the-shelf components. For example, a low-end spoofer is as cheap as \$223~\cite{zeng2018all}, and higher-end ones that can simultaneously track 10+ satellites and transmit 10+ fake GPS signals only cost similar to a laptop~\cite{nighswander2012gps, utaustinspoofer}. Considering such high realism, in this paper we consider it as a practical attack vector to AD localization.

%% file: threat_model.tex
\vspace{-0.1in}
\nsection{Attack Model and Problem Formulation}\label{sec:threat_model}
\vspace{0.07in}

\nsubsection{Attack Goal and Incentives} \label{sec:attack_goal}
\vspace{-0.05in}

\begin{table*}
\footnotesize
    \begin{minipage}{0.66\linewidth}
		\centering
        \includegraphics[width=0.9\linewidth]{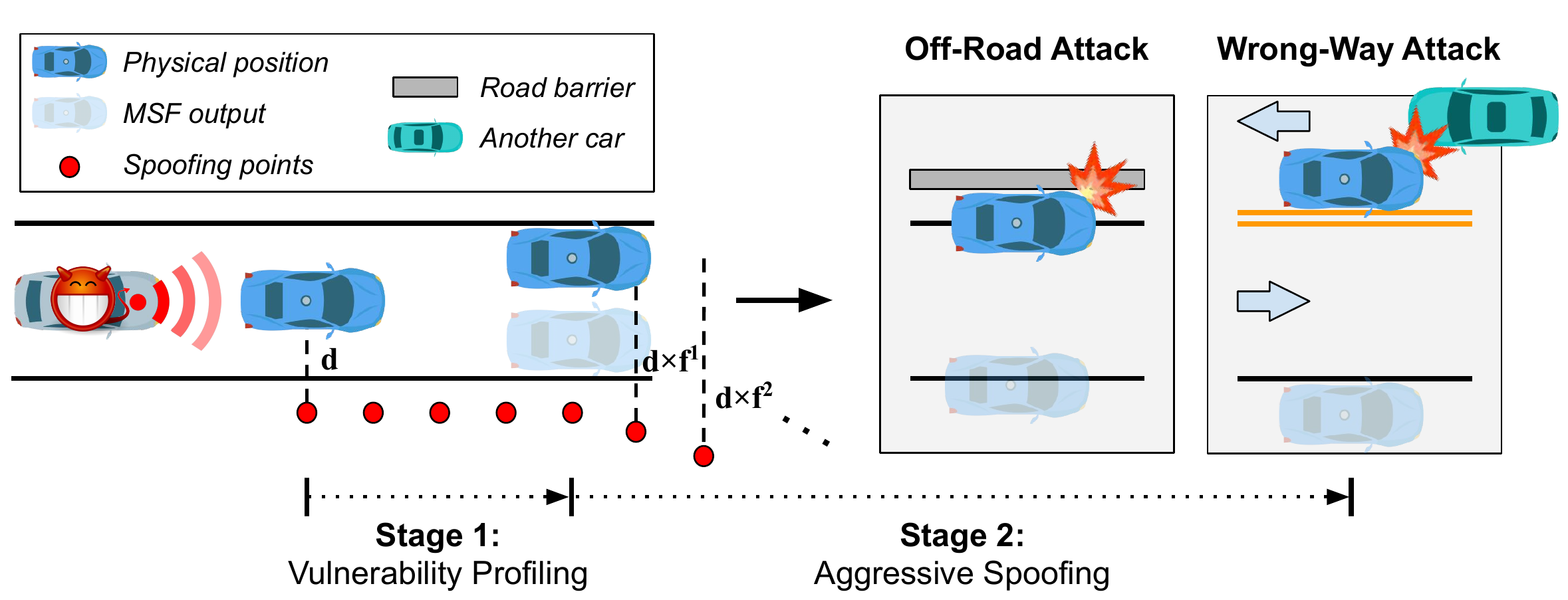}
        \vspace{-0.15in}
		\captionof{figure}{Illustration of the 2-stage attack design and consequences of FusionRipper.}
		\label{fig:goals_and_scenarios}
	\end{minipage}\hfill
	\begin{minipage}{0.31\linewidth}
        \ncaption{Required deviations for the two attack goals considered in this paper. The values are calculated based on common AV, lane, and road shoulder widths (detailed in Appendix~\ref{appendix:goal_deviations}).}
        \label{tbl:goal_deviations}
        \vspace{-0.1in}
		\centering
        \begin{tabular}{@{}ccc@{}}
        \toprule
        \multirow{2}{*}{Attack Goal} & \multicolumn{2}{c}{Required Deviation (m)} \\ \cmidrule(l){2-3} 
         & Local & Highway \\ \midrule
        Off-Road Attack & 0.895 & 1.945 \\
        Wrong-Way Attack & 2.405 & 2.855 \\ \bottomrule
        \end{tabular}
	\end{minipage}
	\vspace{-0.15in}
\end{table*}

\textbf{Attack goals}. In this paper, we target an attack scenario where an attack vehicle tailgates a victim AV while launching a GPS spoofing attack, which is both practical and effective as evaluated by previous work using real cars~\cite{zeng2018all}.
In such a scenario, we consider an attack goal of introducing large \textit{lateral} deviations to the localization output of the victim AV, \ie, deviating to the left or right. Since all vehicles need to drive within their designated road lanes for safety protections, such lateral deviations can pose a direct threat to road safety.

In particular, in this paper we consider two concrete attack goals specific to the AV context: \textit{off-road attacks} and \textit{wrong-way attack}. As illustrated in Fig.~\ref{fig:goals_and_scenarios}, the former aims at deviating to either left or right until the victim drives off the road pavement, while the latter aims at deviating to the left until the victim drives on the opposite traffic lane. Table~\ref{tbl:goal_deviations} lists the required deviations to achieve these two goals, which will be used in our subsequent security analysis.

In the AV context, these two attack goals can cause various \textit{safety hazards specific to localization errors} such as driving off road to hit road curbs or falling down the highway cliff. Since in high-level AD systems the perception module is only designed for obstacle detection and the localization module is in full charge of identifying road deviations~\cite{udacity_av_nd, udacity_av_apollo, coursera_av, apollo, autoware}, these hazards cannot be prevented even when the perception module is functioning perfectly.
Moreover, such hazards cannot be prevented even if high-level AD systems directly use perception sensors, e.g., cameras and ultrasonic sensors, for collision avoidance.
These two attack goals can also cause \textit{vehicle collisions}, \eg, with vehicles in adjacent or opposite traffic lanes. Even when the AV can perform automatic emergency brake, it cannot avoid being hit by other vehicles that fail to yield on time, especially those human driving ones with over 2 sec average driver reaction time~\cite{cali_driver_reaction_time}.

\textbf{Attack incentives.} No matter whether road accidents are caused, the victim AVs under the two attack goals are already violating the traffic rules~\cite{offroad_violate_law, wrongway_violate_law} and exhibiting unsafe driving behaviors. These can already damage the reputation of the corresponding AV company. Thus, a likely attack incentive is \textit{business competition}, which can allow one AV company to deliberately damage the reputation of its rival AV companies and thus unfairly gain competitive advantages. This is especially realistic today considering that there are over 40 companies competing in the AV market~\cite{av-companies}. Meanwhile, considering the direct safety impact, we also cannot rule out the possible incentives for terrorist attacks or targeted murders, \eg, against civilians, or controversial politicians or celebrities.

\vspace{-0.07in}
\nsubsection{Threat Model} \label{sec:threat_model_details}
\vspace{-0.05in}

\textbf{Attacker's capability.} We assume that the attacker can launch GPS spoofing (\S\ref{sec:spoofing}) to control the positioning measurements of the victim's GPS receiver, with a similar level of measurement uncertainty as the natural GPS signals.
We also assume that the attacker can track the physical positions of the victim AV in the real time during the tailgating. This can be achieved by computing the attack vehicle's own position and offsetting it with the relative position between the attack vehicle and the victim. One concrete scenario is that the attack vehicle is also an AV with a similar set of sensors and run state-of-the-art AD localization algorithms for its own position and AD perception algorithms for the relative position.
Under this scenario, the attacker can thus accurately track the victim positions since for AVs precisely tracking the positions of surrounding obstacles in the real time is one of the most basic tasks for ensuring correct and safe driving.
Such a scenario is especially realistic when the attack is from rival AV companies (incentive discussed in \S\ref{sec:attack_goal}).

\textbf{AV control assumption.} We assume that AD systems are designed to drive on the center of traffic lanes, and constantly tries to correct any deviation to the center. State-of-the-art AD systems from both the academia~\cite{paden2016survey} and industry~\cite{apollo,autoware} follow such design and use lateral controllers to enforce it at a high frequency in the control module (\eg, 100 Hz in Apollo~\cite{apollo}). This means that when the attacker introduces a deviation to the MSF output (\eg, to the right in Fig.~\ref{fig:goals_and_scenarios}), the victim AV will actively correct it and thus cause its physical-world position to have the same amount of deviation but to the \textit{opposite} direction (\eg, to the left in Fig.~\ref{fig:goals_and_scenarios}).

\vspace{-0.05in}
\nsubsection{Attack Formulation} \label{sec:formulation}
\vspace{-0.05in}

Based on the attack model above, the attack in our study can be formulated as the following optimization problem:
\vspace{-0.05in}
\begin{equation}
\small
\begin{aligned}
    \underset{\{\delta_k^a \rvert k=1,...,n\}}{\text{max}} \quad & \mathcal{D}(x_{n}^a, \{x_{k} \rvert k=1,...,n\}) \\
    \text{where} \quad & x_k^a=\mathcal{M}(x_{k-1}^a, r_k+\delta_k^a, z^{\text{lidar}}_k, imu_k), x_0^a = x_0,
\end{aligned}
\label{eq:attack_formulation}
\end{equation}
where $\delta_k^a$ is the GPS spoofing distance to the victim's physical-world position $r_k$ on the road plane, $x_k$ is the MSF output without the attack, $x_k^a$ is the MSF output with the attack, $z^{\text{lidar}}_k$ is the LiDAR locator output, $imu_k$ is the IMU measurement, $\mathcal{D}(\cdot)$ denotes the lateral deviation between a position and a trajectory, and $\mathcal{M}(\cdot)$ denotes an iteration in the KF-based MSF algorithm (introduced in~\S\ref{sec:msf}), and $k$ is the iteration index. As shown, mathematically our attack on MSF is to find a sequence of spoofing distances $\{\delta_k^a \rvert k=1,...,n\}$ that can maximize the deviation of the $n$-th attacked MSF output to the original trajectory $\{x_{k} \rvert k=1,...,n\}$.

%% file: security_analysis.tex
\vspace{-0.1in}
\nsection{Security Analysis of MSF Algorithm}\label{sec:security_analysis}
\vspace{-0.07in}

To systematically understand the security property of MSF-based AD localization,
we start with the necessary first step: understanding the upper-bound attack effectiveness, \ie, the maximum possible deviation, under the attack formulation.

\vspace{-0.05in}
\nsubsection{Upper-Bound Attack Effectiveness}\label{sec:exhaustive_search}
\vspace{-0.07in}

\textbf{Analysis methodology.} To analyze the upper-bound attack effectiveness, we perform exhaustive search of possible attack inputs $\{\delta_k^a \rvert k=1,...,n\}$ to the representative MSF implementation, BA-MSF, to find the one that can maximize Eq.~\ref{eq:attack_formulation}.
We did not choose to use an optimizer since the BA-MSF implementation is released in the binary form and thus we cannot directly get its analytical formula.
For a given sensor input trace in our analysis, there are multiple possible \textit{attack windows}, \ie, from one GPS input to another later. For each attack window, we iteratively search for the $\delta_k^a$ that can deviate the most from $x_{k}$, which is a method also used in previous theoretical work on the security of single-source KF~\cite{su2016stealthy,liu2012cyber,mo2010false1,mo2010false2}. In accordance with our threat model, we set the measurement uncertainty of GPS spoofing inputs as the median value in real-world sensor input traces of BA-MSF.

We perform the analysis above on two types of sensor input traces: (1) real-world trace, and (2) synthetic noise-free trace. The former is obtained by directly recording the run-time MSF input while the AV is driving in the real world. Analysis results from this type of traces have the highest realism, but the types of analysis we can perform are limited since we cannot easily modify the sensor data without violating the consistency among different sensor inputs, and the analysis insights can be less clean due to real-world sensor noises. Thus, we complement it with the latter, which synthesizes MSF inputs following a given driving trajectory, with all the LiDAR locator and non-spoofed GPS inputs set to the ground truth positions, their measurement uncertainty set to the medium value in the real-world trace, and the IMU measurements calculated according to the driving trajectory.

\textbf{Experimental setup.} We obtain the official BA-MSF implementation from the Apollo AD system code base~\cite{apollo}. For the real-world trace, we use the BA-MSF input trace released by Apollo, which is recorded in Sunnyvale, CA and 4-min long~\cite{baidudata}. In this paper, we denote it as \textit{ba-local}. For the synthetic trace, we generate one for a common driving trajectory: driving on a straight road with a constant velocity of 45 mph. In our analysis, we use an attack window of 10 attack inputs, which is 10 seconds since the GPS input is 1 Hz in Apollo. In the exhaustive search, we enumerate $\delta_k^a$ from 0 to 10 meters with step size of 0.04 meters on both left and right sides, since we find that in our experiments GPS input deviations larger than that are identified as outliers by the Chi-squared test in BA-MSF. The medium measurement uncertainty values for GPS and LiDAR locator are calculated from trace \textit{ba-local}.

\begin{figure}[tbp]
\centering
\includegraphics[width=.95\columnwidth]{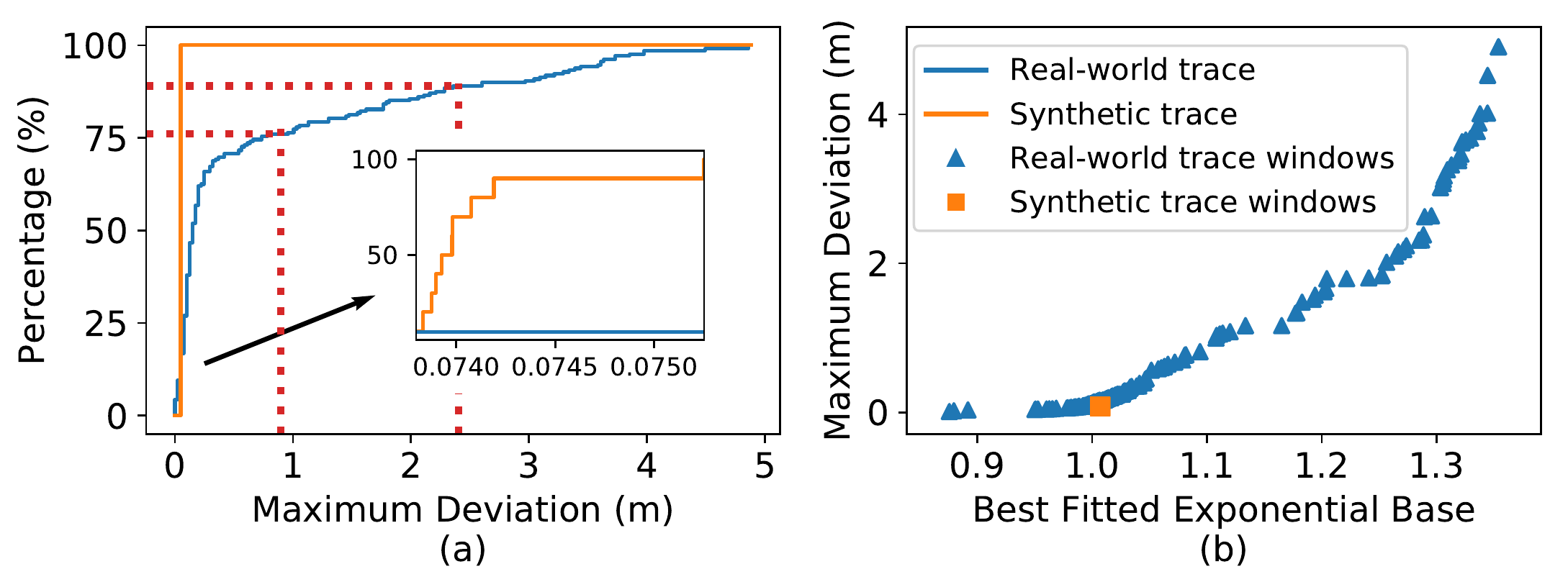}
\vspace{-0.17in}
\ncaption{(a) CDF of the maximum deviations for attack windows in real-world and synthetic traces. Attack goals are marked in red dotted lines. (b) Maximum deviations and best fitted exponential bases of attack windows in the two traces.}
\label{fig:maxdev_expbase}
\vspace{-0.15in}
\end{figure}

\textbf{Results.} Fig.~\ref{fig:maxdev_expbase} (a) shows the distribution of the upper-bound deviations achieved in the 10-point attack windows for each trace. As shown, in both real-world and synthetic traces, even such maximum possible attack effectiveness is very limited: majority (76.0\%) of the attack windows in the real-world trace and \textit{all} of those in the synthetic trace cannot reach even the lowest required deviations (0.895 m) in Table~\ref{tbl:goal_deviations}.
The main reason behind such poor attack performances is as follows.
First, due to outlier detection, the maximum deviation achievable by the first attack input is very small, \eg, at most 0.06 meters. Next, such tiny deviation can be quickly corrected by LiDAR locator inputs since in between two GPS attack inputs there are 5 LiDAR locator inputs (5 Hz in Apollo). This makes it highly difficult for subsequent attack inputs to build upon the deviations achieved by previous attack input. Thus, production-grade KF-based MSF algorithms today can indeed generally enhance the security against GPS spoofing.

At the same time, we also observe that the results between the real-world trace and the synthetic trace have very sharp differences: in the synthetic trace, the upper-bound deviations for all attack windows are at most 0.076 meters, while those in the real-world trace is generally larger, with 90.3\% of them larger than 0.076 meters. This suggests that \textit{sensor noises in the real world can generally degrade the security of MSF}. As shown later, such real-world factors can actually enable highly effective attacks that fundamentally break MSF in practice.

\begin{figure}[tbp]
\centering
\includegraphics[width=.95\columnwidth]{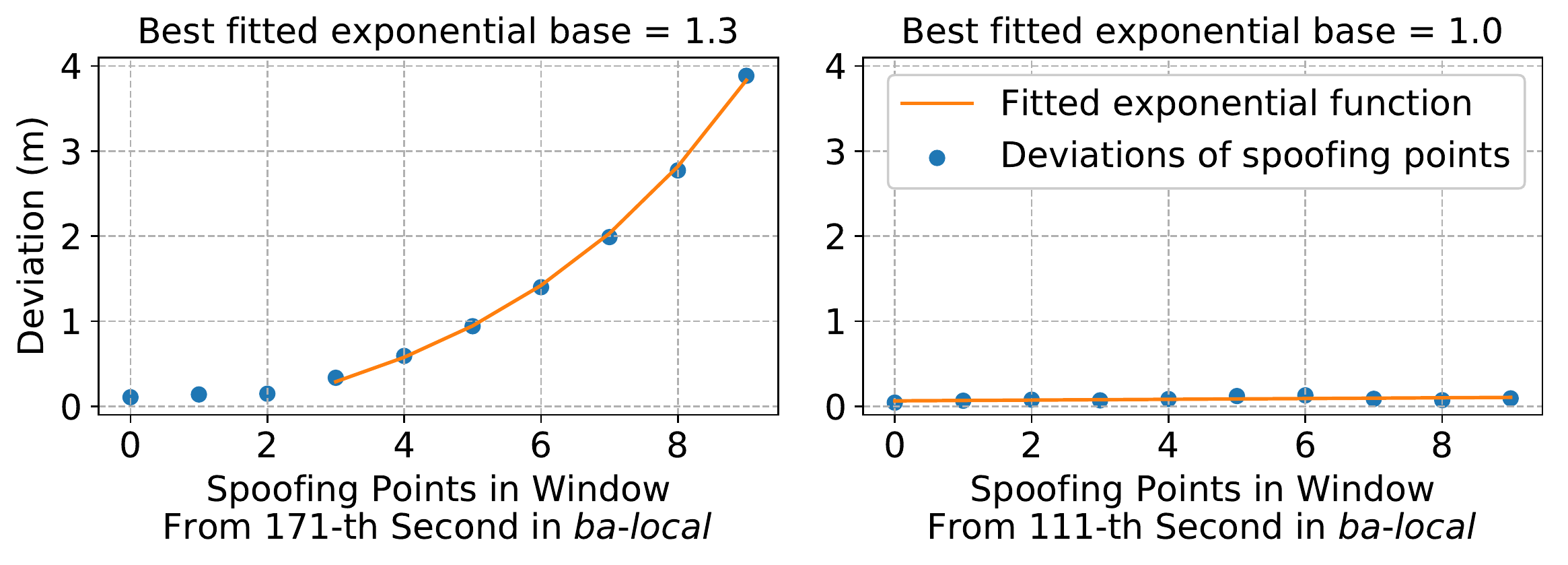}
\vspace{-0.15in}
\ncaption{The deviations and best fitted exponential bases of two example attack windows in the real-world trace. Left is with take-over effect; Right is without take-over effect.}
\label{fig:win_dev_eg}
\vspace{-0.15in}
\end{figure}

\textbf{Observation: take-over effect.} While our results show a general lack of attack capability to achieve even the easiest attack goal in Table~\ref{tbl:goal_deviations}, we also observe that for the real-world trace there still exist 14\% attack windows that can actually achieve over 2 meters deviations, which are large enough for some of our attack goals. For all of these windows, we find that GPS spoofing is able to cause \textit{an exponential growth} of deviations, and one such example is shown on the left of Fig.~\ref{fig:win_dev_eg}. As shown, its deviation trend is very different from those in majority of other attack windows as shown on the right of Fig.~\ref{fig:win_dev_eg}, which is almost flat.

To more quantitatively measure such observation, for each window, we fit an exponential function $f(x)=a^x + b$ to the deviations, where $x$ is the $x$-th attack point and $f(x)$ is the deviations. For each 10-point window, we use the exponential base $a$ in the best fitted function (based on the mean squared error) to measure the exponential growth trend. As shown in Fig.~\ref{fig:maxdev_expbase} (b), such exponential growth trends have strict positive correlation with the upper-bound deviations in the attack windows, and all windows that can have very large deviations, \eg, over 3 meters for achieving \textit{all} attack goals in Table~\ref{tbl:goal_deviations}, have very clear exponential growth trend, \eg, with $a$ being at least 1.3 (the trend on the left of Fig.~\ref{fig:win_dev_eg}).

\begin{figure}[tbp]
\centering
\includegraphics[width=.52\columnwidth]{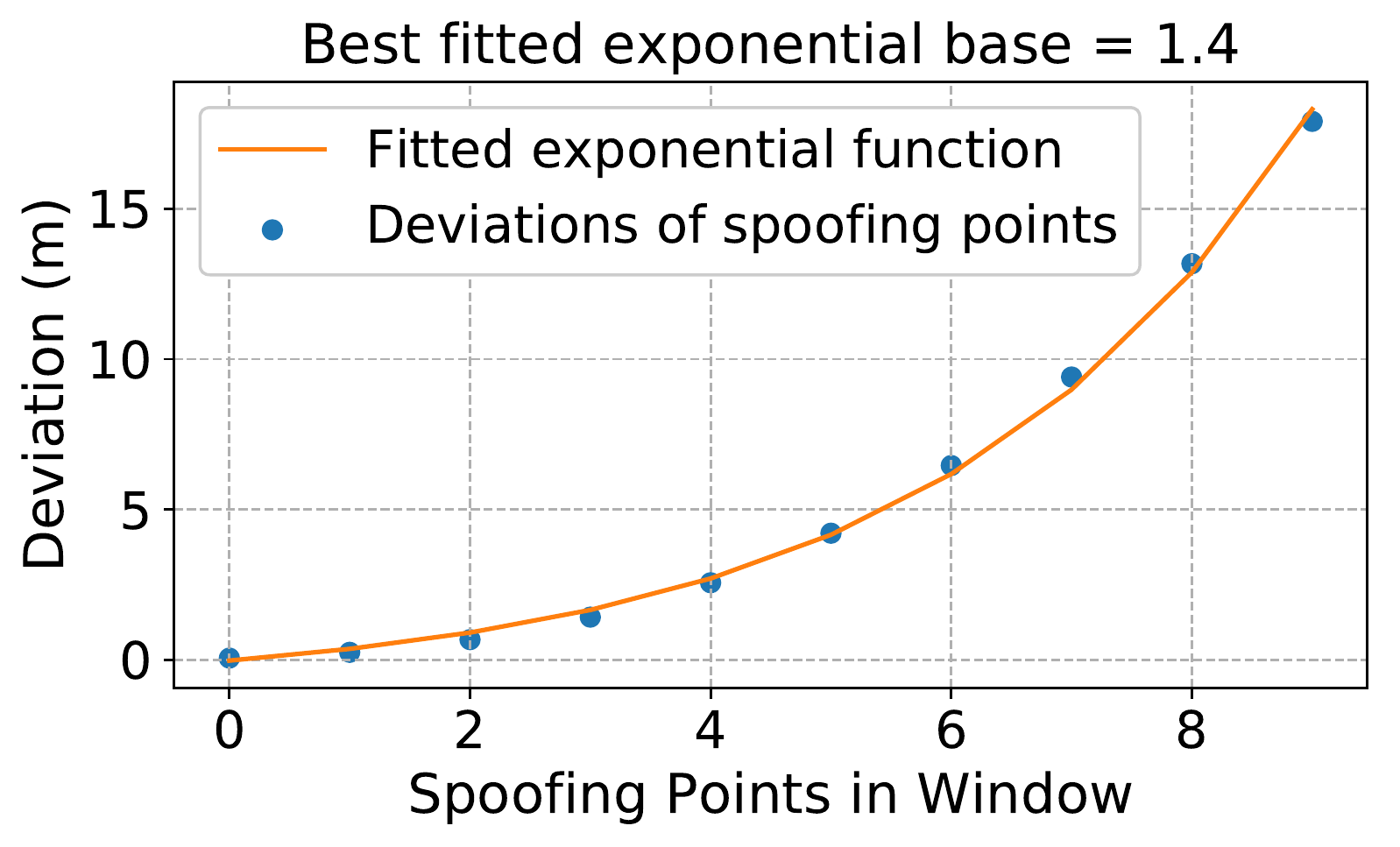}
\vspace{-0.15in}
\ncaption{The deviation growth and the best fitted exponential base for BA-MSF with only the spoofed GPS input in KF updates (or a \textit{single-source} KF-based MSF) in the synthetic trace under exhaustive search.}
\label{fig:single_source}
\vspace{-0.25in}
\end{figure}

Such exponential growth trend is very similar to the situation when the spoofed GPS is the only positioning source in KF updates, which is confirmed by re-running the upper-bound attack analysis in the synthetic trace without LiDAR locator inputs as shown in Fig.~\ref{fig:single_source}.
This means that for these windows with exponential deviation growths, GPS inputs somehow become the dominating KF update source (we will analyze the cause later). In fact, according to the Chi-squared test values in the analysis logs, we find that LiDAR locator inputs actually become outliers in the latter parts of these windows and then can not provide corrections any more. This thus \textit{fundamentally defeats the design principle of MSF, \ie, the fusion of multiple input sources for more robustness and accuracy}. In this paper, we call it \textit{take-over effect}.

For an attacker, such take-over effect is the most desired attack outcome, since it can efficiently cause \textit{arbitrary deviations} and thus lead to both off-road and wrong-way attacks, and even larger ones if desired. Thus, in the next section we perform a cause analysis to understand why such take-over effect appears in the real-world trace.

\vspace{-0.05in}
\nsubsection{Cause Analysis} \label{sec:cause_analysis}
\vspace{-0.07in}

Since take-over effect does not appear in all attack windows, there must be some factors other than the attack input $\delta^a_k$ that contribute to the take-over opportunity. To analyze the causes for take-over effect, we first identify possible contributing factors using theoretical analysis and experimental validation, and then use correlation analysis to identify the most important factors for the observed take-over effect in our analysis.

\begin{figure}[tbp]
\centering
\includegraphics[width=\columnwidth]{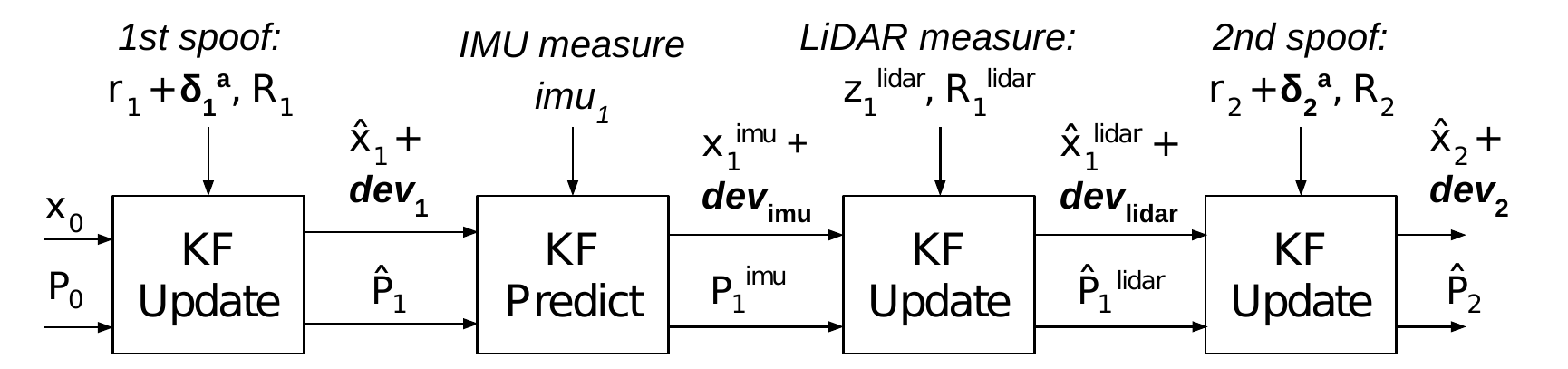}
\vspace{-0.22in}
\ncaption{A simplified but general MSF operation pipeline under GPS spoofing attack for theoretical analysis.}
\label{fig:equation_flow}
\vspace{-0.1in}
\end{figure}

\textbf{Derivation of contributing factors.}
To identify the set of possible contributing factors to the deviations in MSF, we first perform theoretical analysis based on the general KF-based MSF design (\S\ref{sec:msf}).
For the ease of the theoretical analysis without loss of generality, we target the smallest unit in the attack, the MSF operation pipeline between two consecutive GPS spoofing inputs, and simplify it to only have one IMU input and one LiDAR locator input. Fig.~\ref{fig:equation_flow} shows such simplified pipeline, and the notations we use in the analysis, where $dev_{1}$, $dev_{\text{imu}}$, $dev_{\text{lidar}}$, and $dev_{2}$ denote the MSF output deviations after the first GPS spoofing input, the IMU input, the LiDAR locator input, and the second GPS spoofing inputs.

Here we derive the deviations after each step in the simplified but general KF-based MSF operation pipeline for theoretical contributing factor analysis. Table~\ref{tbl:notations} (in Appendix) lists the math notations and their descriptions used in the derivation.

First, after spoofing the 1st GPS point with a spoofing distance $\delta_1^a$, the KF state equations become:
\vspace{-0.05in}
\begin{equation*}
\small
\begin{split}
    \mathbf{\hat{x}}_{1}^a &= \mathbf{x}_0 + \mathbf K_1(\mathbf r_1 + \delta_1^a - \mathbf H \mathbf x_0) \\
    &= \mathbf{\hat{x}}_1 + \mathbf K_1 \delta_1^a \\
    \mathbf{\hat{P}}_1 &= \mathbf{P}_0 - \mathbf K_1 \mathbf H \mathbf{P}_0 \\
    \mathbf K_{1} &= \mathbf P_0 \mathbf H^T(\mathbf H \mathbf P_0 \mathbf H^T + \mathbf R_{1})^{-1} \\
\end{split}
\end{equation*}

Thus, the deviation after spoofing the first point is:
\vspace{-0.05in}
\begin{equation*}
\small
\begin{split}
    dev_{1} &= \mathbf K_1 \delta_1^a \\
\end{split}
\end{equation*}

Second, we perform an IMU prediction. IMU values are used in the kinematics function described by the matrix $\mathbf F_1$:
\vspace{-0.05in}
\begin{equation*}
\small
\begin{split}
    \mathbf x_1^{\text{imu},a} &= \mathbf F_1 \mathbf{\hat{x}}_{1}^a \\
    &= \mathbf F_1 (\mathbf{\hat{x}}_1 + \mathbf K_1 \delta_1^a) \\
    &= \mathbf x_1^{\text{imu}} + \mathbf F_1 \mathbf K_1 \delta_1^a \\
    \mathbf P_1^{\text{imu}} &= \mathbf F_1 \mathbf{\hat{P}}_{1} \mathbf F_1^T + \mathbf Q \\
\end{split}
\end{equation*}

After the IMU prediction, the deviation becomes:
\vspace{-0.05in}
\begin{equation*}
\small
\begin{split}
    dev_{\text{imu}} &= \mathbf F_1 dev_{1} \\
\end{split}
\end{equation*}

Third, a LiDAR locator update is applied. $\Delta_{\text{lidar}}=\mathbf x_1^{\text{imu}} - \mathbf z_1^{\text{lidar}}$ describes the distance between the LiDAR position and the original non-spoofed KF state. This is because sensor noises or LiDAR locator inaccuracies will cause LiDAR locator outputs to be misaligned with the MSF output.
\vspace{-0.05in}
\begin{equation*}
\small
\begin{split}
    \mathbf{\hat{x}}_1^{\text{lidar},a} &= \mathbf{x}_1^{\text{imu},a} + \mathbf K_1^{\text{lidar}} (\mathbf z_1^{\text{lidar}} - \mathbf H \mathbf x_1^{\text{imu},a}) \\
    &= \mathbf x_1^{\text{imu}} + dev_{\text{imu}} \\
    & \quad + \mathbf K_1^{\text{lidar}} (\mathbf z_1^{\text{lidar}} - \mathbf H (\mathbf x_1^{\text{imu}} + dev_{\text{imu}})) \\
    &= \mathbf{\hat{x}}_1^{\text{lidar}} + dev_{\text{imu}} - \mathbf K_1^{\text{lidar}} (\Delta_{\text{lidar}} + dev_{\text{imu}}) \\
    \mathbf{\hat{P}}_1^{\text{lidar}} &= \mathbf{P}_1^{\text{imu}} - \mathbf K_1^{\text{lidar}} \mathbf H \mathbf{P}_1^{\text{imu}} \\
    \mathbf K_1^{\text{lidar}} &= \mathbf P_1^{\text{imu}} \mathbf H^T(\mathbf H \mathbf P_1^{\text{imu}} \mathbf H^T + \mathbf R_1^{\text{lidar}})^{-1} \\
\end{split}
\end{equation*}

LiDAR locator output provides correction on the deviation. After the KF update, the deviation then becomes:
\vspace{-0.05in}
\begin{equation*}
\small
\begin{split}
    dev_{\text{lidar}} &= dev_{\text{imu}} - \mathbf K_1^{\text{lidar}} (\Delta_{\text{lidar}} + dev_{\text{imu}}) \\
\end{split}
\end{equation*}

Finally, we spoof the second GPS point with the spoofing distance $\delta_2^a$:
\vspace{-0.05in}
\begin{equation*}
\small
\begin{split}
    \mathbf{\hat{x}}_{2}^a &= \mathbf{\hat{x}}_1^{\text{lidar},a} + \mathbf K_2(\mathbf r_2 + \delta_2^a - \mathbf H \mathbf{\hat{x}}_1^{\text{lidar},a}) \\
    &= \mathbf{\hat{x}}_1^{\text{lidar}} + dev_{\text{lidar}} \\
    & \quad + \mathbf K_2 (\mathbf r_2 + \delta_2^a - \mathbf H (\mathbf{\hat{x}}_1^{\text{lidar}} + dev_{\text{lidar}})) \\
    &= \mathbf{\hat{x}}_2 + dev_{\text{lidar}} + \mathbf K_2 (\delta_2^a - dev_{\text{lidar}}) \\
    \mathbf{\hat{P}}_2 &= \mathbf{\hat{P}}_1^{\text{lidar}} - \mathbf K_2 \mathbf H \mathbf{\hat{P}}_1^{\text{lidar}} \\
    \mathbf K_2 &= \mathbf{\hat{P}}_1^{\text{lidar}} \mathbf H^T(\mathbf H \mathbf{\hat{P}}_1^{\text{lidar}} \mathbf H^T + \mathbf R_{2})^{-1} \\
\end{split}
\end{equation*}

And the deviation after the second spoofing point will be:
\vspace{-0.05in}
\begin{equation*}
\small
\begin{split}
    dev_{2} &=  dev_{\text{lidar}} + \mathbf K_2 (\delta_2^a - dev_{\text{lidar}}) \\
\end{split}
\end{equation*}

Based on the derivation, there are 4 theoretical contributing factors to $dev_{2}$ besides the attack input $\delta_k^a$:
\begin{itemize}
\item \textit{Initial MSF state uncertainty} ($\mathbf P_0$): The larger $\mathbf P_0$ is, the less confident the MSF algorithm has on its positioning output, and thus more updates are taken from attack inputs $\delta_k^a$, leading to larger $dev_{2}$.
\item \textit{LiDAR measurement uncertainty} ($\mathbf R^{\text{lidar}}_1$): The larger $R^{\text{lidar}}_1$ is, the less confident the LiDAR locator is on its positioning output $z^{\text{lidar}}_1$, and thus the larger the \textit{remaining} deviation after LiDAR locator's correction $dev_{\text{lidar}}$, leading to larger $dev_{2}$.
\item \textit{Difference between LiDAR position and the original MSF output without attack} ($\Delta_{\text{lidar}}$): The impact of $\Delta_{\text{lidar}}$ on $dev_{2}$ has two phases. First, as $\Delta_{\text{lidar}}$ increases, the correction from the LiDAR update increases, which causes $dev_{\text{lidar}}$ to be smaller and decreases $dev_{2}$. Second, after $\Delta_{\text{lidar}}$ becomes too big that makes $z^{\text{lidar}}_1$ an outlier, no correction can be applied any more and thus $dev_{2}$ becomes larger than before. Thus, there is a non-linear relationship between $\Delta_{\text{lidar}}$ and $dev_{2}$.
\item \textit{IMU measurement} ($imu_1$): $imu_1$ affect on $dev_{2}$ in two ways. First, $imu_1$ is used in $\mathbf F_1$ (the IMU-based integration function in Eq.~\ref{eq:kf_pred}), which directly affects $dev_{\text{imu}}$ and further affects $dev_{2}$. Second, $\mathbf F_1$ affects $\mathbf P_{\text{imu}}$ and then the Kalman gain at LiDAR update $\mathbf K_{\text{lidar}}$ and at the second spoofing $\mathbf K_{2}$ (Eq.~\ref{eq:kf_update}). Note that larger $\mathbf K_{\text{lidar}}$ means larger correction and thus smaller $dev_{2}$, while larger $\mathbf K_{2}$ means larger $dev_{2}$. Thus, the relationship between $imu_1$ and $dev_{2}$ depends on the design of $\mathbf F_1$ and the competition of the impact of larger $\mathbf K_{\text{lidar}}$ and larger $\mathbf K_{2}$ on $dev_{2}$.
\end{itemize}

\begin{figure}[tbp]
\centering
\includegraphics[width=.75\columnwidth]{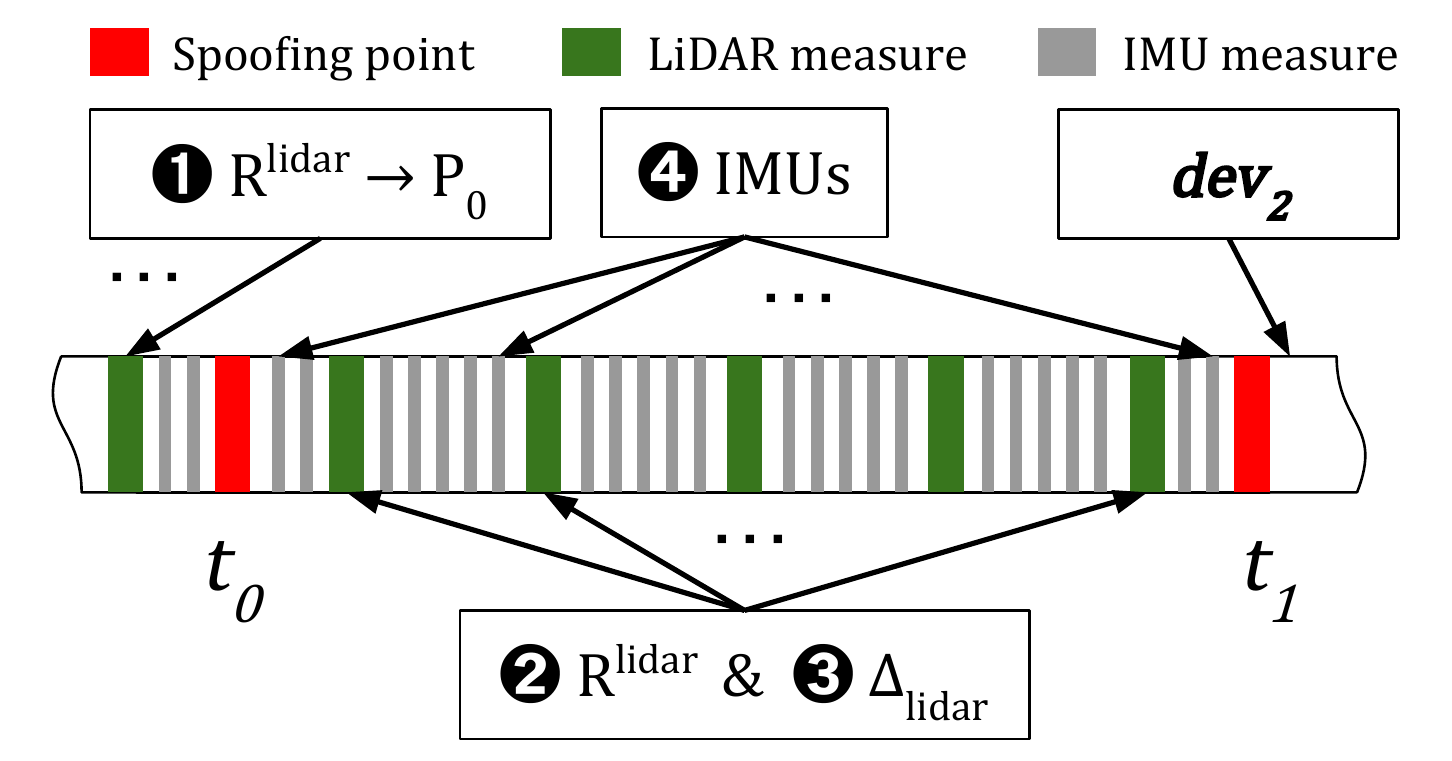}
\vspace{-0.1in}
\ncaption{Modeling of each factor in the synthetic trace. We modify different parts of the sensor data in order to observe how the factors affect the 2nd deviation $dev_{2}$.}
\label{fig:factor_modeling}
\vspace{-0.15in}
\end{figure}

\textbf{Experimental validation of derived contributing factors.}
To validate whether these 4 factors indeed affect the actual BA-MSF implementation, we take a segment from the synthetic sensor trace as shown in Fig.~\ref{fig:factor_modeling}, and modify different parts of sensor data to model the change of the four contributing factors. As shown, the segment consists of two GPS spoofing points. Since no spoofing has been applied prior to time $t_0$, the deviation prior to $t_0$ is zero. Unlike the simplified MSF operation pipeline considered in the theoretical derivation, we apply the original KF prediction and update sequences as real-world sensor traces for BA-MSF, \ie, 1 Hz for GPS, 5 Hz for LiDAR locator, and 200 Hz for IMU~\cite{apollo}.

For each contributing factor, we measure the deviation after the 2nd spoofing point to understand the its relationship with the deviation. To eliminate the influence from the GPS spoofing distance, we exhaustively search for different distances for two GPS spoofing points and use the best one in our results. We use the median value of the GPS uncertainty in \textit{ba-local} as the uncertainty values for the GPS spoofing points, which is the same as in \S\ref{sec:threat_model}.

\begin{figure}[tbp]
\centering
\includegraphics[width=.85\columnwidth]{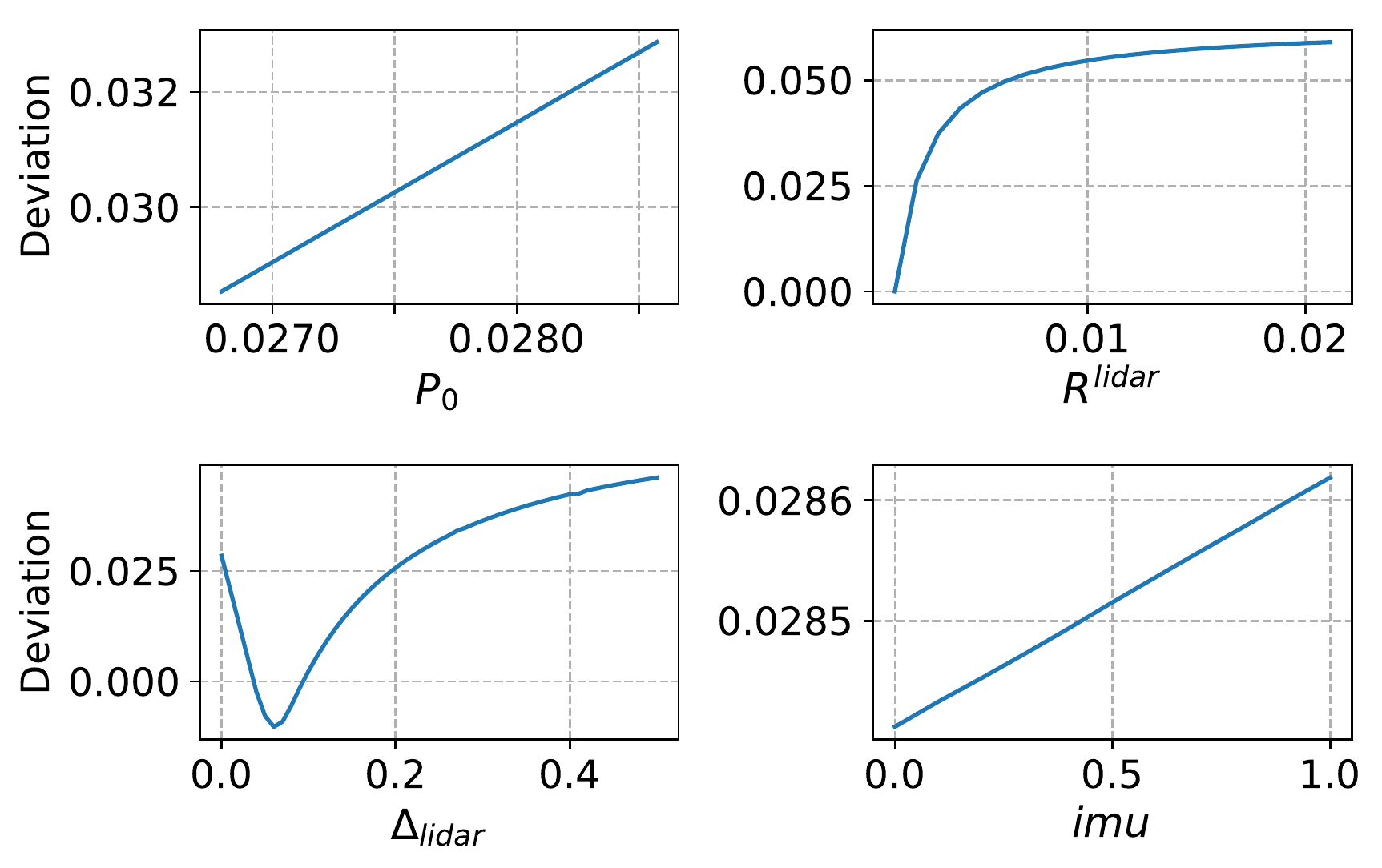}
\vspace{-0.1in}
\ncaption{Relationship between the contributing factors and the spoofing deviation in the synthetic trace.}
\label{fig:modeling_result}
\vspace{-0.2in}
\end{figure}

\textbf{Validation results.} The experiment results are shown in Fig.~\ref{fig:modeling_result} and described below:
\begin{itemize}
\item For $\mathbf P_0$, there is no direct way to modify it since it is not part of sensor data. Here we vary $\mathbf R^{\text{lidar}}$ before $t_0$ to \textit{indirectly} generate different values of $\mathbf P_0$. Since the LiDAR locator outputs are aligned with the MSF state (\ie, $\Delta_{\text{lidar}} = 0$) before $t_0$, the change of $\mathbf R^{\text{lidar}}$ will only affect $\mathbf P_0$. As shown in the top-left subfigure in Fig.~\ref{fig:modeling_result}, the results validate that a larger $\mathbf P_0$ will cause a larger deviation $dev_2$.
\item We modify $\mathbf R^{\text{lidar}}$ between time $t_0$ and $t_1$ to observe how it affects the correction capability of LiDAR locator outputs on the deviation. As shown in the top-right subfigure in Fig.~\ref{fig:modeling_result}, our results validate that $\mathbf R^{\text{lidar}}$ has a positive effect on the deviation, but reaches a plateau when it is overly large.
\item The modeling of $\Delta_{\text{lidar}}$ is straightforward: We directly set the LiDAR locator outputs to have different distances to the MSF state.
The bottom-left subfigure in Fig.~\ref{fig:modeling_result} shows our results. As shown, aligned with our theoretical analysis, a small $\Delta_{\text{lidar}}$ can correct the deviation introduced by the first GPS spoofing. However, when $\Delta_{\text{lidar}}$ increases, at a certain point it causes the MSF output to deviate to the opposite direction. This is because $\Delta_{\text{lidar}}$ provides a large update to the velocity component in the MSF state such that the deviation is over-corrected as time accumulates. When $\Delta_{\text{lidar}}$ becomes even larger, the deviation starts to increase since LiDAR locator outputs become outliers, which also conforms to our theoretical analysis.
\item For $imu$, we modify the acceleration component in the IMU measurements between time $t_0$ and $t_1$. In addition, we align the LiDAR locator positions to the \textit{non-spoofed} MSF outputs during this period to ensure that $\Delta_{\text{lidar}}=0$. As shown in the bottom-right subfigure in Fig.~\ref{fig:modeling_result}, our results show that $imu$ has a positive influence on the deviation overall, which shows that the impact of $\mathbf K_{2}$ is much larger than the impact of $\mathbf K_{\text{lidar}}$ to $dev_{2}$.
\end{itemize}

\textbf{Factor importance analysis}. With the 4 contributing factors identified, we then use popular causality analysis methods to understand the importance of these factors on causing the take-over effect observed in~\S\ref{sec:exhaustive_search}. Specifically, we perform the exponentiation function fitting as described in~\S\ref{sec:exhaustive_search}, and label the windows with exponential base $a$ over 1.1 as windows with take-over effect. As shown in Fig.~\ref{fig:maxdev_expbase} (b), for windows without any take-over effect, \eg, the ones for the synthetic trace, the exponential base $a$ is way below 1.1. 
With the exponential fitting results, we identify the first point of the exponential growth to obtain $\mathbf P_0$. For $\mathbf R^{\text{lidar}}$, $\Delta_{\text{lidar}}$, and $imu$, we use the average values from the first point of the exponential growth to the end of the window. We use 2 statistical testing methods commonly used for causality analysis~\cite{medeiros2017software, pajouh2016two, brown2013understanding}: Pearson’s Correlation and Fisher’s Exact Test.

\textbf{Analysis results.} Table~\ref{tbl:factor_importance} shows the experiment results. For the two statistical testing methods, $p < 0.05$ is considered statistically significant, and $r > 0.5$ and $or > 9$ are considered strongly correlated for Pearson's Correlation and Fisher's Exact Test respectively~\cite{cohen2013statistical}. As shown, only the $p$ values for $\mathbf P_0$ and $\mathbf R^{\text{lidar}}$ are statistically significant for both methods, with their $r$ values very close to showing strong correlations, and their $or$ values showing strong correlations. In contrast, neither of the $r$ or $or$ values for $\Delta_{\text{lidar}}$ and $imu$ show strong correlations, and for $imu$, the results are not even statistically significant. This suggests that the take-over effect we observe in our upper-bound analysis is most likely caused by relatively large $\mathbf P_0$ and $\mathbf R^{\text{lidar}}$ in the corresponding attack windows.

For these two most important contributing factors, $\mathbf R^{\text{lidar}}$ reflects the lack of confidence in the LiDAR-based localization algorithm during the attack window, and $\mathbf P_0$ reflects the lack of confidence in the KF states at the beginning of the attack window. This means that \textit{take-over opportunities, or vulnerabilities, appear when the MSF is in relatively unconfident periods}. Because of this, the MSF algorithm needs to take more updates from the GPS inputs, the relatively most confident input source in that period,
which thus allows GPS inputs to dominate KF updates
and trigger the take-over effect.

Since $\mathbf R^{\text{lidar}}$ is the uncertainty reported by LiDAR locator, a large $\mathbf R^{\text{lidar}}$ is caused by the inaccuracies of such locator algorithm in practice. From 
the KF equations (\S\ref{sec:msf}), a large $\mathbf P_0$ is mainly caused by larger uncertainties from the LiDAR locator and GPS updates before the attack window, which is thus due to algorithm inaccuracies in LiDAR locator and noises in GPS signals.
Thus, unconfident periods in MSF are mainly created by practical factors such as algorithm inaccuracies and sensor noises. This also explains why we cannot observe any take-over effect in synthetic noise-free trace. 
These practical factors are fundamentally difficult to avoid in practice, which is exactly why MSF is designed to compensate such inaccuracies and noises from individual sources~\cite{udacity_av_nd,wan2018robust,gao2015ins,suhr2016sensor,tao2013mapping,schreiber2016vehicle,de2017survey,soloviev2008tight,lee2015gps,kelly2011visual}. However, as shown in our analysis, \textit{even for the high-end sensors used in AVs today, these inaccuracies and noises are unfortunately large and frequent enough for GPS spoofing to exploit and fundamentally break MSF in practice}.

\begin{table}[tbp]
\footnotesize
\centering
\ncaption{Correlations between the contributing factors and the take-over vulnerability. Results with statistically strong correlation are highlighted in bold.}
\vspace{-0.1in}
\label{tbl:factor_importance}
\setlength{\tabcolsep}{3.2pt}
\begin{tabular}{@{}ccccc@{}}
\toprule
\multirow{2}{*}{\begin{tabular}[c]{@{}c@{}}Correlation\\ Method\end{tabular}} & \multicolumn{4}{c}{Factor Importance} \\ \cmidrule(l){2-5} 
 & $\mathbf P_0$ & $\mathbf R^{\text{lidar}}$ & $\Delta_{\text{lidar}}$ & $imu$ \\ \midrule
\begin{tabular}[c]{@{}c@{}}Pearson's\\ Correlation\end{tabular} & \textbf{0.42 (2.0e-10)} & \textbf{0.44 (3.5e-11)} & 0.12 (8.4e-2) & 0.01 (8.6e-1) \\
\begin{tabular}[c]{@{}c@{}}Fisher's\\ Exact Test\end{tabular} & \textbf{21.09 (8.6e-6)} & \textbf{11.78 (5.2e-8)} & 5.91 (3.2e-4) & 1.95 (1.1e-1) \\ \bottomrule
\multicolumn{5}{l}{Pearson's correlation: $r$ ($p$-value), where r is the correlation coefficient} \\
\multicolumn{5}{l}{Fisher's exact test: $or$ ($p$-value), where $or$ is the odds ratio}
\end{tabular}
\vspace{-0.25in}
\end{table}

%% file: attack_design.tex
\vspace{-0.05in}
\nsection{Attack Design: FusionRipper}\label{sec:attack_design}
\vspace{-0.07in}

Although our analysis in \S\ref{sec:security_analysis} reveals that there do exist take-over vulnerabilities for MSF in the real world, such vulnerabilities only appear in the unconfident periods created by dynamic and non-deterministic practical factors such as algorithm inaccuracies and sensor noises, which is not observable by the attacker in a tailgating attack vehicle (\S\ref{sec:threat_model}) and are highly difficult, if not impossible, to directly control. Thus, the attacker has to \textit{opportunistically} capture and exploit such vulnerable periods in the actual attack time.

Leveraging this idea, we propose a novel attack design against MSF-based AD localization, called \textit{FusionRipper}, which consists of 2 stages as depicted in Fig.~\ref{fig:goals_and_scenarios}:

\textbf{Stage 1: Vulnerability profiling.} In this stage, the attacker performs GPS spoofing and measures the feedback from the victim AV to profile when vulnerable periods appear. In our design, we aim for as fewer attack parameters as possible to maximize the ease of implementation and robustness, and thus choose to use \textit{constant spoofing} for this stage, \ie, always setting $\delta_k^a$ to a constant $d$ as shown in Fig.~\ref{fig:goals_and_scenarios}. Although such profiling method is simple, our evaluation results later in \S\ref{sec:evaluation} show that it is able to achieve a high attack success rate that is very close to the theoretical upper bound.

While performing constant spoofing, the attacker tracks victim's physical positions in real time and measures their deviations to the center of traffic lane (described in~\S\ref{sec:threat_model}). If such deviation is as large as causing the AV to exhibit unsafe driving behaviors, \eg, about to have unnecessary lane straddling, the victim AV is considered as in the vulnerable period. Our design uses the deviation that can touch the left or right lane line on local roads (0.295 meters, detailed in Appendix~\ref{appendix:goal_deviations}) as the threshold to determine vulnerable periods. The intuition is that a properly designed and tested AD system 
should very rarely have large position deviations that can cause unsafe driving behaviors under normal fluctuations of sensor inputs. For example, the errors of BA-MSF evaluated by Baidu Apollo AVs on real roads are within 0.054 meters~\cite{wan2018robust}, which is far less than 0.295 meters. Thus, when such rare deviation appears, it is very likely caused by the constant spoofing, and the MSF algorithm is very likely in an unconfident period since it takes larger update from the spoofed GPS inputs.

\textbf{Stage 2: Aggressive spoofing.} After the vulnerable period is identified, the attacker can then perform aggressive spoofing to trigger the take-over effect and thus quickly induce large deviations. As shown in our security analysis in~\S\ref{sec:exhaustive_search}, the deviations grow exponentially during the take-over effect, and thus we choose exponential spoofing in the aggressive spoofing stage. As shown in Fig.~\ref{fig:goals_and_scenarios}, as soon as the attacker identifies a vulnerable period, she switches to use spoofing distance $d \times f^i$, where an exponential base $f$ is cumulatively multiplied to previous spoofing distance at each of the spoofing points, and $i$ is the index of the aggressive spoofing inputs.

\textbf{Generality.} Since FusionRipper is designed to exploit the take-over vulnerability that is general to any KF-based MSF as discussed in our cause analysis based on the general form of KF-based MSF (\S\ref{sec:cause_analysis}), its design is generally applicable to any KF-based MSF algorithms. As shown in our generality evaluation later (\S\ref{sec:generality}), FusionRipper is highly effective on different KF-based MSF designs and implementations.

%% file: evaluation.tex
\vspace{-0.1in}
\nsection{Attack Evaluation}\label{sec:evaluation}
\vspace{0.07in}

\nsubsection{Evaluation Methodology} \label{sec:eval_method}
\vspace{-0.05in}

\textbf{Experimental setup.} Following the common practice among AV companies~\cite{frossard2018end, gao2020vectornet}, we evaluate FusionRipper on real-world sensor traces. Specifically, we use
the real-world trace \textit{ba-local} used in our security analysis earlier (\S\ref{sec:security_analysis}), and also traces from KAIST Complex Urban~\cite{jeong2019complex}, a dataset for evaluating AD systems. Since \textit{ba-local} is collected by the Apollo team and is designed specifically for evaluating MSF-based localization algorithms for Apollo, it is by default compatible with BA-MSF with a complete positioning sensor set as well as the HD Map for running the LiDAR locator\footnote{Apollo released 8 sensor traces recorded with localization, but only \textit{ba-local} has both the complete sensor set and compatible format with BA-MSF.}.

Similar to \textit{ba-local}, the traces in the KAIST dataset are also collected by high-end AV-grade positioning sensors~\cite{jeong2019complex}. But unfortunately, they do not provide the HD Map for running the LiDAR locator in BA-MSF. To address this, we assume an \textit{ideal} LiDAR locator which always outputs the ground truth positions provided in the KAIST dataset, with their measurement uncertainty set to the median value of that in \textit{ba-local}. Considering that one of the likely causes for the take-over effect is the LiDAR locator inaccuracies, especially the measurement uncertainty values (\S\ref{sec:cause_analysis}), this assumption only makes the attack harder and thus the results will provide the worst-case attack effectiveness on the KAIST traces.

\textbf{Trace selection in KAIST dataset.}
The KAIST dataset includes 18 local traces and 2 highway traces that are compatible with BA-MSF, and we select 3 local ones and both the 2 highway ones. We truncate them to the first 5 minutes to keep the evaluation time manageable. In the selection of local traces, we select the ones with the smallest average MSF state uncertainty (\ie, most confident).
Table~\ref{tbl:kaisttraces} shows the average MSF state co-variance value, \ie, uncertainty, when running BA-MSF on the 20 traces in the KAIST dataset that (1) have the complete sensor data needed by BA-MSF, \eg, some KAIST traces do not have complete IMU data, and (2) from a stationary position to provide a complete motion history, which is required for BA-MSF to have stable outputs.
Among the 18 local traces and 2 highway traces, we choose both the eligible highway traces, and select the top 3 from the local traces with the lowest MSF state uncertainties. Considering that state uncertainty is one of the two most important contributing factors to the take-over effect (\S\ref{sec:exhaustive_search}), the evaluation results on these traces will provide the worst-case attack effectiveness for the KAIST traces.

\begin{table}[tbp]
\ncaption{Average MSF co-variance, \ie, uncertainty, of the KAIST local and highway traces. We ranked the traces based on their MSF state co-variance (the lower the more confident), and pick the most confident ones (in \textbf{bold}) in our evaluation.}
\label{tbl:kaisttraces}
\vspace{-0.1in}
\begin{minipage}{\linewidth}
\footnotesize
\centering
\setlength{\tabcolsep}{4pt}
\begin{tabular}{@{}cccccc@{}}
\toprule
\begin{tabular}[c]{@{}c@{}}Local\\ Trace\end{tabular} & \begin{tabular}[c]{@{}c@{}}Avg. MSF\\ Co-variance\end{tabular} & Rank & \begin{tabular}[c]{@{}c@{}}Local\\ Trace\end{tabular} & \begin{tabular}[c]{@{}c@{}}Avg. MSF\\ Co-variance\end{tabular} & Rank \\ \midrule
\textbf{\textit{ka-local08}} & \textbf{0.0032} & \textbf{1} & \textit{ka-local39} & 0.1143 & 10 \\
\textbf{\textit{ka-local31}} & \textbf{0.0080} & \textbf{2} & \textit{ka-local16} & 0.2254 & 11 \\
\textbf{\textit{ka-local07}} & \textbf{0.0111} & \textbf{3} & \textit{ka-local29} & 0.3237 & 12 \\
\textit{ka-local37} & 0.0131 & 4 & \textit{ka-local09} & 0.4070 & 13 \\
\textit{ka-local35} & 0.0146 & 5 & \textit{ka-local14} & 0.4468 & 14 \\
\textit{ka-local33} & 0.0219 & 6 & \textit{ka-local38} & 0.8904 & 15 \\
\textit{ka-local36} & 0.0312 & 7 & \textit{ka-local26} & 1.4719 & 16 \\
\textit{ka-local28} & 0.1026 & 8 & \textit{ka-local27} & 6.4191 & 17 \\
\textit{ka-local30} & 0.1029 & 9 & \textit{ka-local3}2 & 33.3712 & 18 \\ \bottomrule \\
\end{tabular}
\end{minipage}\vspace{-0.05in}
\begin{minipage}{\linewidth}
\footnotesize
\centering
\begin{tabular}{@{}ccc@{}}
\toprule
Highway Trace & Avg. MSF Co-variance & Rank \\ \midrule
\textbf{\textit{ka-highway17}} & \textbf{0.0027} & \textbf{1} \\
\textbf{\textit{ka-highway06}} & \textbf{0.0028} & \textbf{2} \\ \bottomrule
\end{tabular}
\end{minipage} 
\vspace{-0.1in}
\end{table}

\textbf{Evaluation metrics.} To evaluate the attack effectiveness, we apply attack parameters $d$ and $f$  from all possible attack starting points, \ie, when the GPS input comes, in each trace, since the attacker can discover the victim at any moment in the trace and start performing the attack. As described earlier in~\S\ref{sec:attack_design}, the attacker switches to aggressive spoofing when the lateral deviation between the spoofed MSF output and the non-spoofed MSF output is over 0.295 meters, which is just about to have lane straddling on local roads.

We consider the attack as successful when the lateral deviation of the MSF output is over the required deviations for the off-road and wrong-way attacks according to Table~\ref{tbl:goal_deviations}.
This follows our AD control assumption (\S\ref{sec:threat_model}), which can directly considers the amount of deviation at the MSF output level as the amount of physical position deviations in the opposite direction to the center line. Later in~\S\ref{sec:end_to_end}, we will concretely evaluate this assumption using an end-to-end evaluation with the AD control taking effect.
The \textit{success rate} is calculated as the fraction of the successful attack starting points out of all starting points. For each attack starting point, we enumerate the combinations of $d$ from 0.3 to 2.0 meters, with step size 0.1 meters, and $f$ from 1.1 to 2.0, with step size 0.1. We choose these ranges because we do not find the values out of these ranges can improve the attack effectiveness in our experiments. 
Each $d$ and $f$ combination is then applied to both the left and right side of the driving direction, since both sides are valid for achieving off-road attack (detailed in~\S\ref{sec:attack_goal}).
Since it takes time to (1) capture a take-over vulnerability, which is created dynamically and non-deterministically, and (2) reach the required deviations even during take-over effects (\S\ref{sec:exhaustive_search}), we also consider \textit{minimum attack duration} when calculating success rate, \ie, how much time the attack can last when tailgating the victim AV. Intuitively, the longer such duration is, the higher chance she can have to hit a vulnerable period.

\begin{table*}[tbp]
\footnotesize
	\begin{minipage}{0.38\linewidth}
        \centering
        \ncaption{Real-world sensor traces used in our evaluation.}
        \label{tbl:datasets}
        \vspace{-0.1in}
        \setlength{\tabcolsep}{3.2pt}
        \begin{tabular}{@{}ccccc@{}}
        \toprule
        Source & Trace Label & Road Type & Duration & HD Map \\ \midrule
        Apollo & \textit{ba-local} & Local & 257s & Yes \\ \midrule
        \multirow{5}{*}{\begin{tabular}[c]{@{}c@{}}KAIST\\ Complex\\ Urban\end{tabular}} & \textit{ka-local08} & Local & 289s & \multirow{5}{*}{No} \\
         & \textit{ka-local31} & Local & 1014s &  \\
         & \textit{ka-local07} & Local & 553s &  \\
         & \textit{ka-highway17} & Highway & 1186s &  \\
         & \textit{ka-highway06} & Highway & 1937s &  \\ \bottomrule
        \end{tabular}
	\end{minipage}\hfill
    \begin{minipage}{0.6\linewidth}
		\centering
        \includegraphics[width=.92\columnwidth]{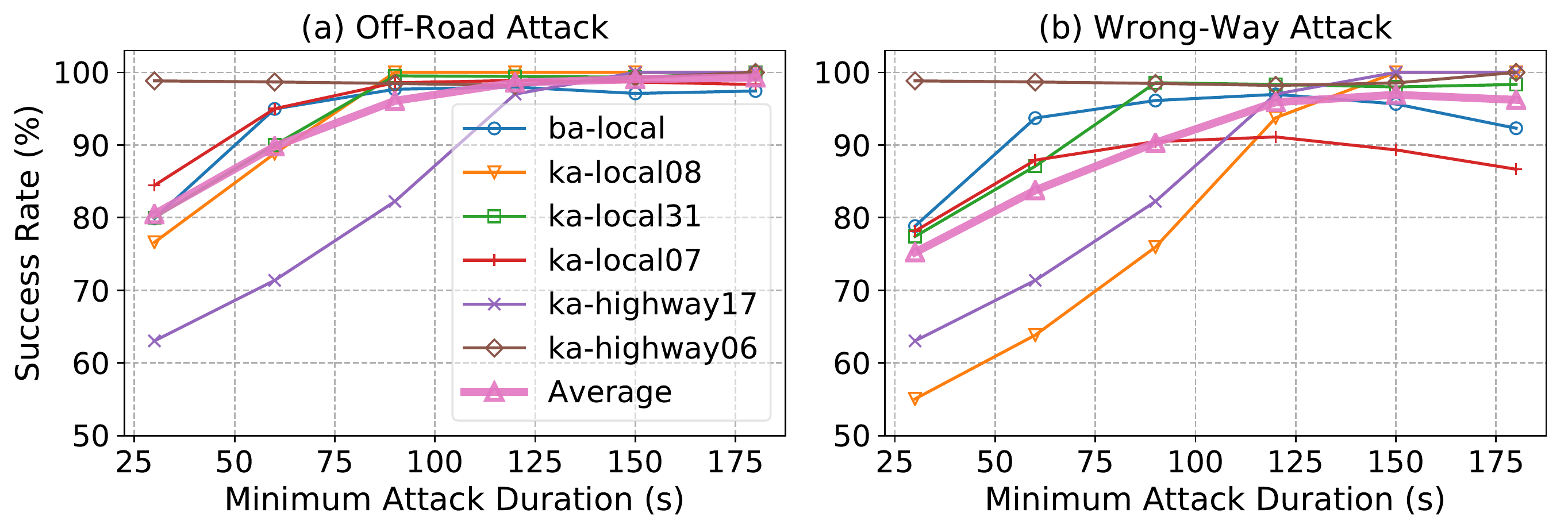}
        \vspace{-0.15in}
        \captionof{figure}{Average attack success rates of (a) \textit{off-road} attack and (b) \textit{wrong-way} attack under different minimum attack duration.}
        \label{fig:single_succrate}
	\end{minipage}
	\vspace{-0.1in}
\end{table*}

\begin{table*}[tbp]
\footnotesize
	\begin{minipage}{0.38\linewidth}
        \centering
        \ncaption{Ablation study results on \textit{ba-local} trace.}
        \label{tbl:lowerbound}
        \vspace{-0.1in}
        \setlength{\tabcolsep}{4pt}
        \begin{tabular}{@{}ccccc@{}}
        \toprule
        \multirow{2}{*}{Attack Config.} & \multicolumn{2}{c}{Off-Road} & \multicolumn{2}{c}{Wrong-Way} \\ \cmidrule(l){2-5} 
         & \begin{tabular}[c]{@{}c@{}}Succ.\\ Rate\end{tabular} & \begin{tabular}[c]{@{}c@{}}Succ.\\ Time\end{tabular} & \begin{tabular}[c]{@{}c@{}}Succ.\\ Rate\end{tabular} & \begin{tabular}[c]{@{}c@{}}Succ.\\ Time\end{tabular} \\ \midrule
        FusionRipper & 98.0\% & 29s & 97.0\% & 33s \\ \midrule
        \begin{tabular}[c]{@{}c@{}}Vulnerability Profiling\\ Stage Only\end{tabular} & 14.1\% & 26s & 7.0\% & 29s \\ \midrule
        \begin{tabular}[c]{@{}c@{}}Aggressive Spoofing\\ Stage Only\end{tabular} & 10.1\% & 8s & 5.0\% & 13s \\ \bottomrule
        \end{tabular}
	\end{minipage}\hfill
    \begin{minipage}{0.29\linewidth}
		\centering
        \includegraphics[width=.95\columnwidth]{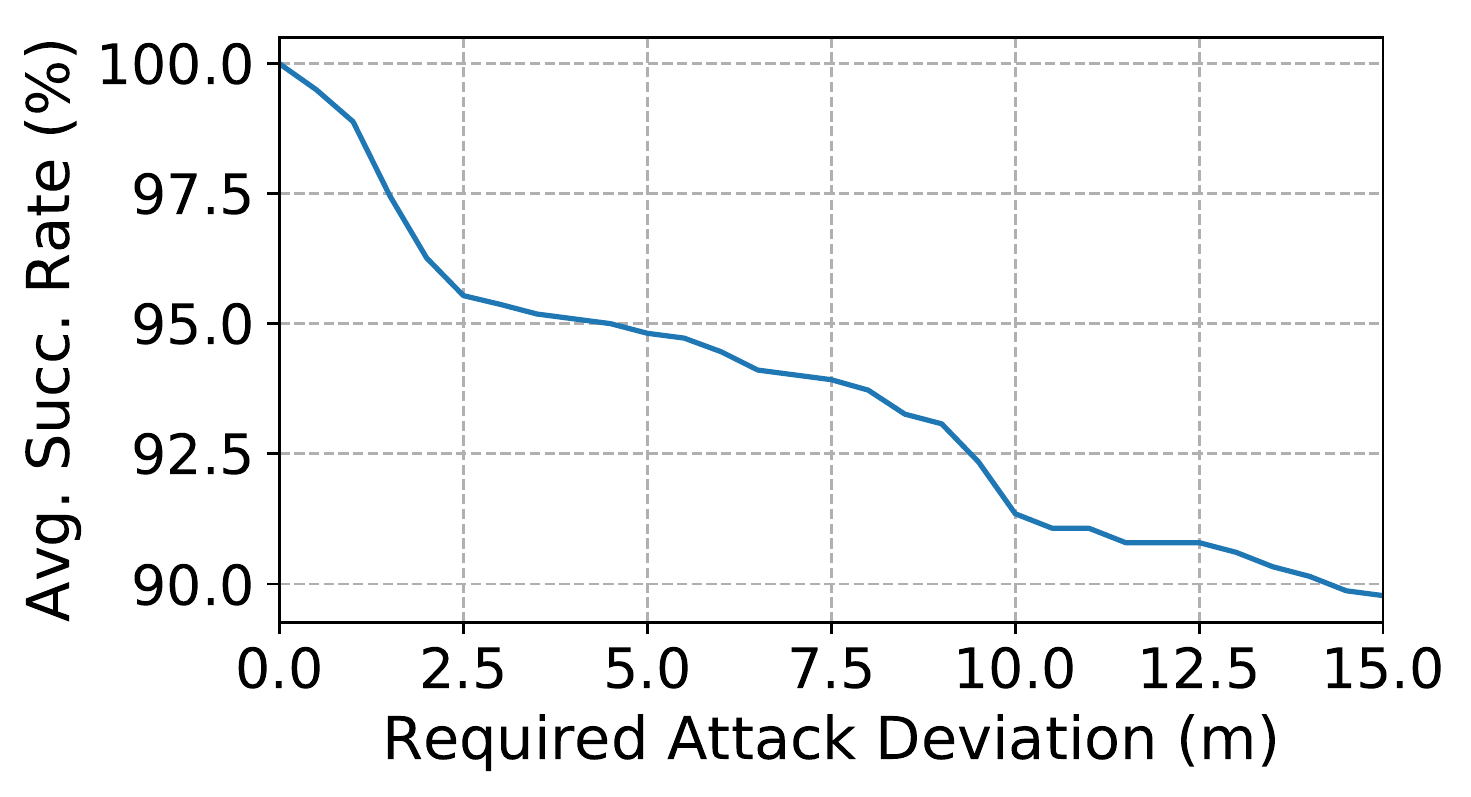}
        \vspace{-0.15in}
        \captionof{figure}{Average success rate under different required attack deviations when the minimum attack duration is 2 minutes.}
        \label{fig:goaldev_succrate}
	\end{minipage}\hfill
	\begin{minipage}{0.29\linewidth}
		\centering
        \includegraphics[width=.95\columnwidth]{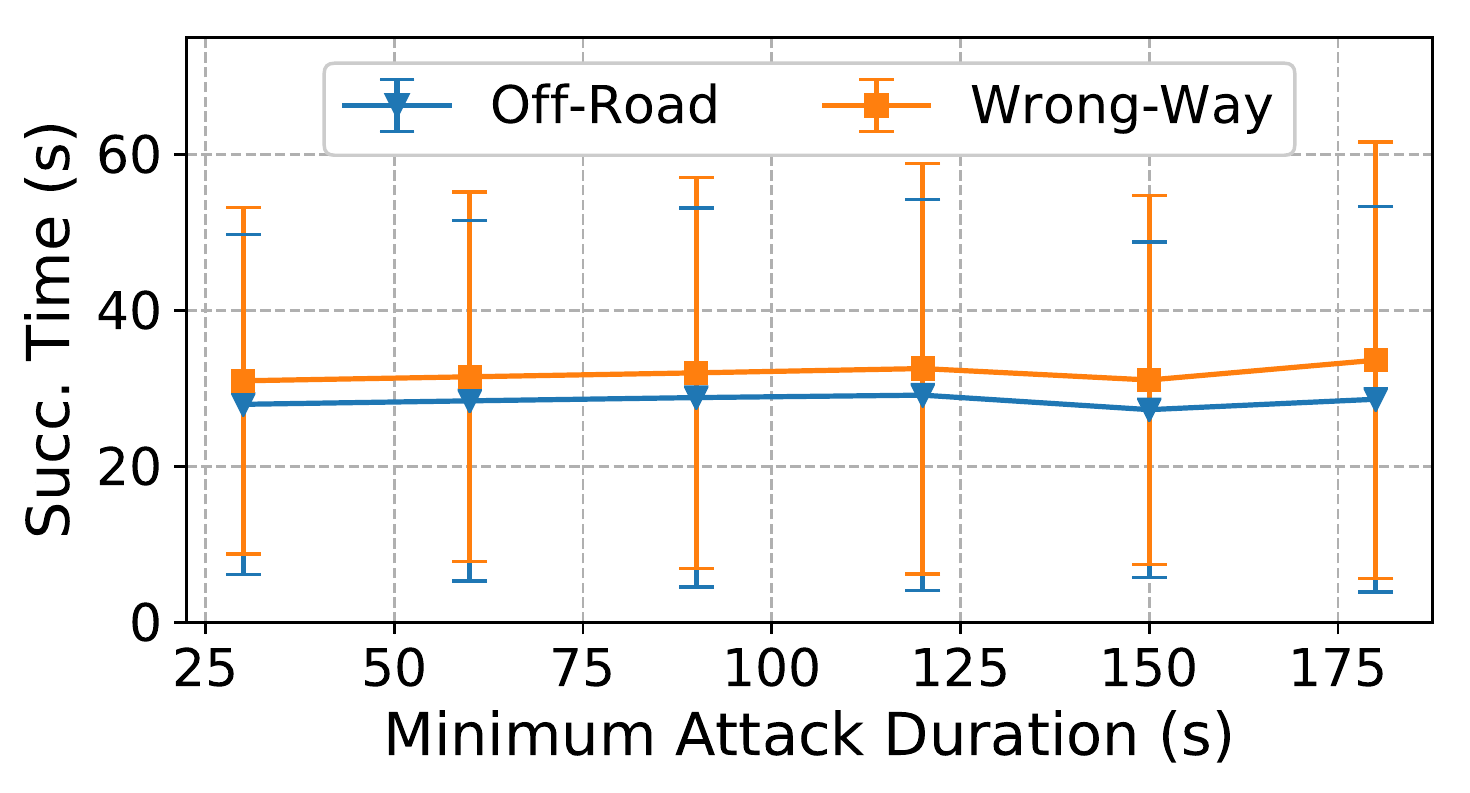}
        \vspace{-0.15in}
        \captionof{figure}{Average success time for reaching required deviations in off-road and wrong-way attacks under different minimum attack duration.}
        \label{fig:single_succtime}
	\end{minipage}
	\vspace{-0.2in}
\end{table*}

\vspace{-0.05in}
\nsubsection{Attack Effectiveness}\label{sec:effectiveness}
\vspace{-0.05in}

\textbf{Attack success rates.} Fig.~\ref{fig:single_succrate} shows the best success rates of FusionRipper among all the combinations of $d$ and $f$ for the two attack goals. It shows both the results for individual traces and the average result among all traces (the thick pink line). As shown, for all traces, the average success rate is always over 75\% for both attack goals even when the minimum attack duration is as low as 30 seconds. When the minimum attack duration increases, the success rates for all traces increase accordingly, which is expected since the attacker has higher chance to capture a vulnerable period. In particular, when the attack can last 2 minutes, \textit{there exists at least one combination of $d$ and $f$ that can achieve over 97\% success rate (98.6\% on average) for the off-road attack and over 91\% success rate (95.9\% on average) for the wrong-way attack, for all traces in our evaluation}. Note that this is in fact the worst-case results for KAIST traces as discussed in~\S\ref{sec:eval_method}. Since a normal taxi or truck trip is usually at least 10 minutes, it is highly likely that an attacker can find such a 2-minute tailgating opportunity in practice to launch the FusionRipper attack.

Among all the traces, \textit{ka-local08} and \textit{ka-highway17} shows the lowest success rate in general, especially when the required deviation is large.
As shown in Table~\ref{tbl:kaisttraces}, both traces have smallest average MSF state uncertainty in their categories (\ie, local and highway). This means that their MSF outputs have the highest confidence and thus are the most difficult to attack as we expect in~\S\ref{sec:eval_method}. This also confirms that we are evaluating the worst-case attack effectiveness on KAIST traces.

Between the two attack goals, the success rates only slightly drop for wrong-way attack since it has a larger required deviation. 
This means that the majority of the captured vulnerable periods have a successful take-over effect that can be exploited to cause different required deviations. To confirm this, we further evaluate the success rates of FusionRipper for even larger required deviations, and find that when the minimum attack duration is 2 minutes, FusionRipper is able to maintain an average success rate over 91.3\% even when the required deviation is 10 meters as shown in Fig.~\ref{fig:goaldev_succrate}.

\begin{table*}[tbp]
\footnotesize
\centering
\ncaption{Top 3 attack parameters with the highest attack success rates when minimum attack duration is 2 min.}
\label{tbl:top3_df}
\vspace{-0.1in}
\setlength{\tabcolsep}{4pt}
\begin{tabular}{@{}cccccccccccccccccccc@{}}
\toprule
\multirow{2}{*}{Attack} & \multirow{2}{*}{Rank} & \multicolumn{3}{c}{\textit{ba-local}} & \multicolumn{3}{c}{\textit{ka-local08}} & \multicolumn{3}{c}{\textit{ka-local31}} & \multicolumn{3}{c}{\textit{ka-local07}} & \multicolumn{3}{c}{\textit{ka-highway17}} & \multicolumn{3}{c}{\textit{ka-highway06}} \\ \cmidrule(l){3-20} 
 &  & \textit{d} & \textit{f} & \begin{tabular}[c]{@{}c@{}}Succ.\\ Rate\end{tabular} & \textit{d} & \textit{f} & \begin{tabular}[c]{@{}c@{}}Succ.\\ Rate\end{tabular} & \textit{d} & \textit{f} & \begin{tabular}[c]{@{}c@{}}Succ.\\ Rate\end{tabular} & \textit{d} & \textit{f} & \begin{tabular}[c]{@{}c@{}}Succ.\\ Rate\end{tabular} & \textit{d} & \textit{f} & \begin{tabular}[c]{@{}c@{}}Succ.\\ Rate\end{tabular} & \textit{d} & \textit{f} & \begin{tabular}[c]{@{}c@{}}Succ.\\ Rate\end{tabular} \\ \midrule
\multirow{3}{*}{Off-Road} & Top 1 & 0.6 & 1.5 & 98.0\% & 0.7 & 1.1 & 100\% & 0.5 & 1.2 & 99.4\% & 0.3 & 1.1 & 98.9\% & 0.3 & 1.2 & 97.0\% & 1.1 & 1.5 & 98.2\% \\
 & Top 2 & 0.6 & 1.6 & 98.0\% & 0.7 & 1.2 & 100\% & 1.0 & 1.3 & 99.4\% & 0.3 & 1.2 & 98.3\% & 0.3 & 1.3 & 97.0\% & 1.1 & 1.3 & 98.2\% \\
 & Top 3 & 0.6 & 1.7 & 98.0\% & 0.7 & 1.3 & 100\% & 1.0 & 1.4 & 99.4\% & 0.4 & 1.2 & 98.3\% & 0.3 & 1.4 & 94.0\% & 1.3 & 1.3 & 98.2\% \\ \midrule
\multirow{3}{*}{Wrong-Way} & Top 1 & 0.6 & 1.5 & 97.0\% & 0.3 & 1.2 & 93.8\% & 1.0 & 1.3 & 98.3\% & 0.3 & 1.4 & 91.1\% & 0.3 & 1.2 & 97.0\% & 1.2 & 1.3 & 98.2\% \\
 & Top 2 & 0.6 & 1.3 & 95.0\% & 0.3 & 1.3 & 93.8\% & 1.0 & 1.2 & 97.8\% & 0.3 & 1.5 & 90.6\% & 0.3 & 1.3 & 97.0\% & 1.3 & 1.3 & 98.2\% \\
 & Top 3 & 0.6 & 1.4 & 95.0\% & 0.5 & 1.3 & 92.1\% & 1.1 & 1.2 & 97.8\% & 0.3 & 1.3 & 88.3\% & 0.3 & 1.4 & 94.0\% & 1.1 & 1.3 & 97.6\% \\ \bottomrule
\end{tabular}
\vspace{-0.2in}
\end{table*}

\textbf{Sensitivity to attack parameters.} Table~\ref{tbl:top3_df}
lists the top 3 combinations for each trace. As shown, the attack effectiveness of FusionRipper is sensitive to the combinations of $d$ and $f$. For example, the best $d$ and $f$ combinations are \textit{all different} for the 6 traces. This motivates us to design an offline method to identify effective $d$ and $f$ combinations to increase the attack practicality, which is detailed later in~\S\ref{sec:profiling}.

\textbf{Ablation study.} The high attack effectiveness
is a result of the combination of the two attack stages. To concretely understand this, we conduct an ablation study on \textit{ba-local},
where we remove one of the two stages in the experiments. For \textit{Vulnerability Profiling Stage Only}, we apply the constant spoofing distance \textit{d} from each starting point. For \textit{Aggressive Spoofing Stage Only}, we directly scale the spoofing distance using different combinations of \textit{d} and \textit{f} from each starting point. For both configurations, we obtain the \textit{highest} success rates by enumerating $d$ or $f$ in the range specified in~\S\ref{sec:eval_method}.

Table~\ref{tbl:lowerbound} shows the experiment results for \textit{ba-local} when the minimum attack duration is 2 minutes. As shown, both configurations can only achieve at most 14\% and 7\% for the two attack goals, which is far less than 98\% and 97\% by FusionRipper. This means that there are still some very unconfident periods that even stage 1 or stage 2 alone can succeed, but as shown, without the help of each other, the success rate is very limited. This concretely demonstrates the necessity of the current 2-stage design of FusionRipper. Note that FusionRipper has longer attack success time than \textit{Aggressive Spoofing Stage Only} due to the time spent on the vulnerability profiling stage. However, since the current $\sim$30 seconds attack time on average is already quite affordable for a tailgating attacker in practice, such advantage is much less important than the much higher success rates by FusionRipper.

\textbf{Attack success time.} For the attack success time, overall the average success time and the standard deviations are very similar under different minimum attack duration as shown in Fig.~\ref{fig:single_succtime}. When the minimum attack duration is 2 minutes, the average success time is less than 30 seconds with a standard deviation of around 25 seconds for both off-road and wrong-way attacks. This shows that FusionRipper can generally succeed very fast, \eg, within a minute, even when the attacker plans to attack for over 2 minutes.

\vspace{-0.05in}
\nsubsection{Comparison with Naive Attack Method} \label{sec:compare_naive_spoofing}
\vspace{-0.05in}

In this section, we compare FusionRipper with a more naive attack method: \textit{random attack}, which randomly spoofs a deviation within a distance range for each GPS spoofing point.

\textbf{Experimental setup.} We perform experiments by applying FusionRipper and random attack on \textit{ba-local}. In the random attack, we uniformly sample the position deviation between 0 to 10 meters for each spoofing point. The experiments are repeated for 30 trials. In each trial, the spoofing is performed for each attack starting point and on both the left and right. The higher success rate between that of the left and that of the right is taken as the final success rate for each trial.

\textbf{Results.} The first row in Table~\ref{tbl:other_msfs} shows the experiment results when the minimum attack duration is 2 minutes. We find that the random attack can barely reach any large deviation, and as shown, its success rates are as low as 3.7\% and 0.2\% on average for the two attack goals respectively, which are much lower than those from FusionRipper (98.0\% and 97\%).

\vspace{-0.05in}
\nsubsection{Generality of FusionRipper}\label{sec:generality}
\vspace{-0.05in}

In this section, we aim at understanding the generality of FusionRipper by evaluating it on more KF-based MSF implementations. Ideally we hope to find other production-grade implementations for AD systems similar to BA-MSF, but to best of our knowledge, BA-MSF is the only publicly-available one so far. Nevertheless, we still try our best to implement/port and evaluate on two other popular KF-based MSF designs, denoted as \textit{JS-MSF} and \textit{ETH-MSF}, which are both designed for general robotics localization instead of for AVs.

\textbf{Experimental setup.} BA-MSF adopts a Linear KF, the most popular KF design for MSF-based localization (Table~\ref{tbl:msf_survey}). Thus, we follow a popular Linear KF based MSF design published by Joan Sol\`a~\cite{jsmsf} and implement JS-MSF. ETH-MSF~\cite{ethmsf_github} is an open-source project developed by researchers from ETH Z\"urich for drones~\cite{ethmsf_iros13}, which implements an Extended KF based MSF, the second popular KF design for MSF-based localization (Table~\ref{tbl:msf_survey}). It has received over 500 stars on GitHub, which is the \textit{highest} among the repositories under the search keyword ``kalman filter sensor fusion''. Both implementations use a Chi-squared test based outlier detector and directly reject outlier measurements. We follow a common parameter tuning process~\cite{groves2015principles} and reach at most 1.91 and 1.17 meters localization accuracies on \textit{ba-local} for JS-MSF and ETH-MSF respectively. Although such accuracies are far from the centimeter-level accuracy required by AD systems, they are common for general robotics localization~\cite{zuo2019visual, zuo2019lic, brossard2018unscented}.

\textbf{Results.} Table~\ref{tbl:other_msfs} shows the attack success rates of FusionRipper and random attack on \textit{ba-local} for all 3 KF-based MSF implementations. As shown, FusionRipper can generally achieve high success rates on all three MSFs, which are 100\% on both JS-MSF and ETH-MSF for both attack goals. However, we also notice that even random attack can also achieve over 95\% success rates for the off-road attack, and over 70\% for the wrong-way attack. This suggests that JS-MSF and ETH-MSF are both very unstable, which can also be seen by the fact that their natural localization errors are already 1.17 and 1.91 meters. In contrast, BA-MSF can achieve 0.054 meters accuracy, which is likely due to additional design features such as zero-velocity update~\cite{wan2018robust}, and better parameter tuning by professional AV engineers. Thus, while our results show that FusionRipper is general for all 3 KF-based MSF implementations, we believe that the results on BA-MSF can more representatively indicate the security status of production-grade MSF-based AD localization today.

\begin{table}[tbp]
\footnotesize
\centering
\ncaption{Attack success rates of FusionRipper and random attack on 3 MSF implementations. The attacks are evaluated on \textit{ba-local} with 2-minute minimum attack duration.}
\label{tbl:other_msfs}
\vspace{-0.1in}
\setlength{\tabcolsep}{4.5pt}
\begin{tabular}{@{}ccccc@{}}
\toprule
\multirow{2}{*}{\begin{tabular}[c]{@{}c@{}}Attacked\\ MSF\end{tabular}} & \multicolumn{2}{c}{FusionRipper} & \multicolumn{2}{c}{Random Attack (avg. of 30 trials)} \\ \cmidrule(l){2-5} 
 & Off-Road & Wrong-Way & Off-Road & Wrong-Way \\ \midrule
BA-MSF & 98.0\% & 97.0\% & 3.7\% & 0.2\%  \\
JS-MSF & 100\% & 100\% & 97.4\% & 92.4\% \\
ETH-MSF & 100\% & 100\%\dag & 95.9\% & 72.5\% \\ \bottomrule
\multicolumn{5}{l}{\dag Achieves 100\% success rate when using a smaller \textit{f} (1.02).}
\end{tabular}
\vspace{-0.3in}
\end{table}

%% file: practical_attack.tex
\vspace{-0.05in}
\nsection{Practical Attack Considerations}\label{sec:practical_attack}
\vspace{-0.07in}

Although FusionRipper already shows very high effectiveness in~\S\ref{sec:evaluation}, we haven't considered two factors that may affect the attack effectiveness in practice: (1) the variations in the spoofed positions and their measurement uncertainty at the victim's GPS receiver, and (2) sensor input changes due to AD control during the attack. In this section, we evaluate the robustness of FusionRipper under these two practical factors. The experiments in this section are mainly performed on the \textit{ba-local} trace since it has the complete set of real-world sensor inputs for BA-MSF and thus has the highest realism.

\vspace{-0.05in}
\nsubsection{Robustness Against Spoofing Inaccuracies}\label{sec:spoofing_inaccuracy}
\vspace{-0.05in}

In~\S\ref{sec:evaluation}, we directly set spoofed GPS inputs $r_k + \delta^a_k$ based on $d$ and $f$, and set their uncertainty $R_k$ as the medium value in real-world traces. However, in practice both can have  variations due to sensor noises. In this section, we denote the variances to $r_k + \delta^a_k$ as $\sigma_{\text{pos}}$, and those to $R_k$ as $\sigma_{\text{var}}$.

\textbf{Inaccuracy sources and modeling.} As specified in our threat model (\S\ref{sec:threat_model}), we assume that the attacker can estimate the victim AV's real-time positions based on her own position and the distance to the victim. Thus, there are three possible error sources for $\sigma_{\text{pos}}$: 1) localization error $\sigma_1$ in attacker's self-localization process, 2) distance measurement error $\sigma_2$ in the measured distance between the attack vehicle and the victim AV, and 3) GPS receiver error $\sigma_3$, \ie, the difference between the position the attacker intended to set and the actual received position at the victim side. Assuming the attacker is equipped with the same sensor set used in an AD system and can run an MSF algorithm of similar quality, $\sigma_1$ will be similar to the inaccuracies of BA-MSF algorithm, which is reported as 0.054 meters in~\cite{wan2018robust}. Since LiDAR can be used to measure the distance to the victim, $\sigma_2$ is thus the distance measurement error in the LiDAR sensor, which is 0.02 meters as specified in the datasheet according to the LiDAR model used in Apollo~\cite{lidardatasheet}. For $\sigma_3$, we directly use the positioning error, 0.01 meters, as specified in the datasheet of the GPS model used in Apollo~\cite{gpsdatasheet}. Assuming that these errors are normally distributed with a zero-mean (common practice in robotics~\cite{thrun2005probabilistic}), the combined distribution for $\sigma_{\text{pos}}$ is conforming to $\mathit{N}(0,\,\sigma_1^{2}+\sigma_2^{2}+\sigma_3^{2})$ = $\mathit{N}(0,\,0.058^2)$.
For the measurement uncertainty error $\sigma_{\text{var}}$ during spoofing, we measure the distribution of GPS measurement uncertainty in the \textit{ba-local} trace, and take the standard deviation $\sigma_{\text{var}} = 0.008$.

\textbf{Experimental setup.} We apply these error distributions to the FusionRipper attack in \textit{ba-local} using the best attack parameter in \textit{ba-local} with 2-minute minimum attack duration. For each GPS spoofing input, we randomly sample a position error from $\mathit{N}(0,\,\sigma_{\text{pos}}^{2})$ and the error direction from a uniform distribution between $0$ to $360$ degrees, and apply them to the spoofed input. Similarly, we randomly sample an error value from $\mathit{N}(0,\,\sigma_{\text{var}}^{2})$ and apply it to the measurement uncertainty of each spoofing input. To further explore the impact of these errors, we also apply 2$\times$ and 3$\times$ amounts of the normal error ($\sigma_{\text{pos}}$ and $\sigma_{\text{var}}$), in our evaluation. We repeat the experiment 100 times for each error amount. 

\begin{figure}[tbp]
\centering
\includegraphics[width=.7\columnwidth]{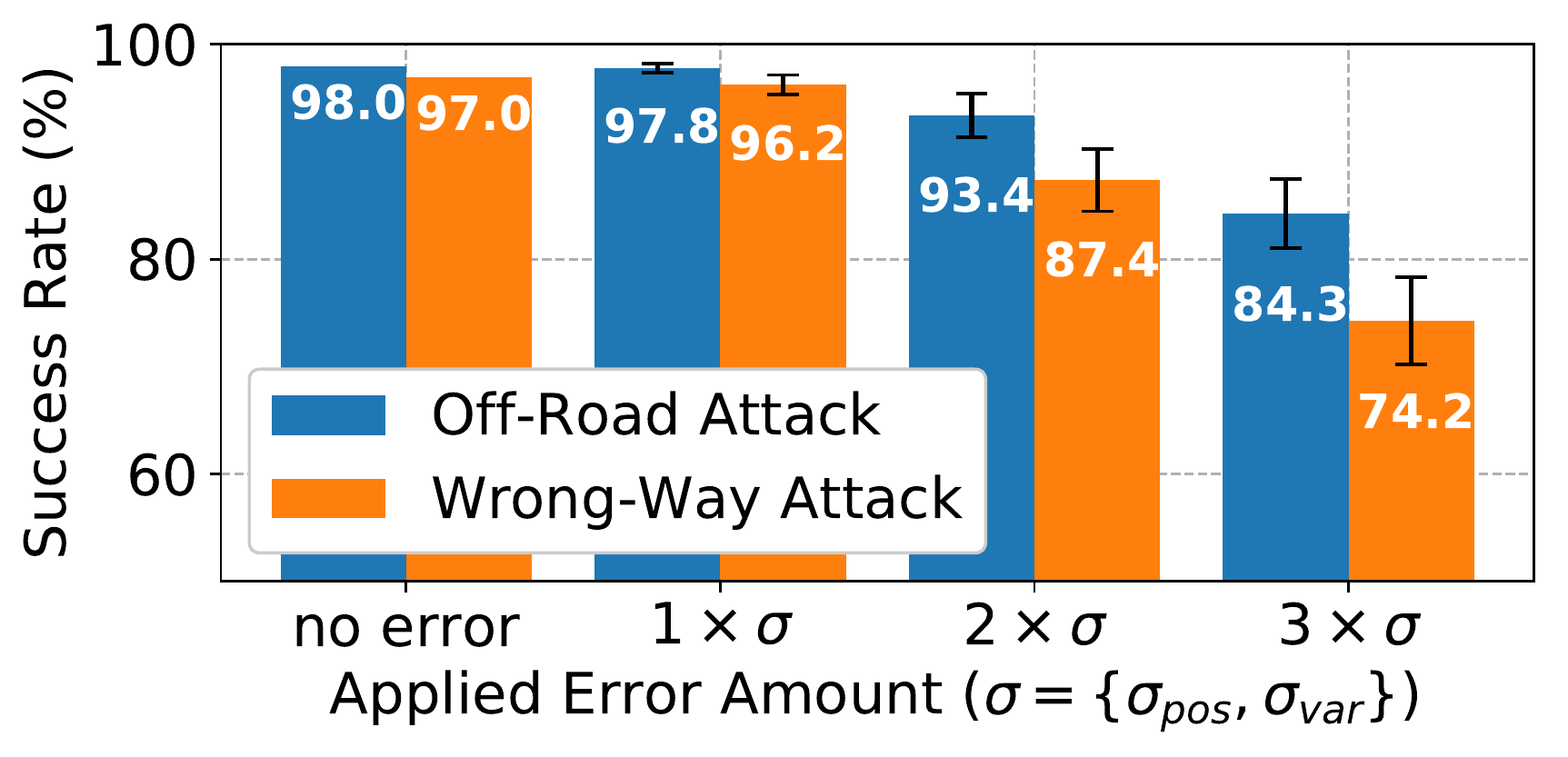}
\vspace{-0.15in}
\ncaption{Attack success rate for different amounts of spoofing errors. Experiment of each error amount is repeated 100 times.}
\label{fig:errors_succrate}
\vspace{-0.25in}
\end{figure}

\textbf{Results.} Fig.~\ref{fig:errors_succrate} shows the attack success rates under each error amount. As shown, under normal error amount ($1\times\{\sigma_{\text{pos}},\sigma_{\text{var}}\}$), the success rate is only reduced by 0.2\% for the off-road attack, and by 0.8\% for the wrong-way attack. Even when the error amount is 3$\times$ than normal, meaning that the error can be as large as 0.174 meters, the success rate is still 84.3\% and 74.2\% on average for off-road and wrong-way attacks respectively. This shows that FusionRipper is highly robust to spoofing inaccuracies in practice.

\vspace{-0.05in}
\nsubsection{End-to-End Attack Impact Evaluation}\label{sec:end_to_end}
\vspace{-0.05in}

In~\S\ref{sec:evaluation}, we assume the amount of deviation in MSF outputs is the same as the amount of physical position deviations to the center line. In this section, we concretely evaluate this assumption by performing an end-to-end attack impact evaluation with the AD control taking effect. %

\textbf{Evaluation methodology.} In this evaluation, we adopt two evaluation methods popularly used in AV industry~\cite{frossard2018end, bansal2018chauffeurnet}: \textit{trace based} and \textit{simulation based}. In the trace-based evaluation, we still use the original real-world sensor trace \textit{ba-local}, and synthesize the sensor input changes corresponding to the output of the control module in Apollo. Specifically, the lateral controller in Apollo runs a linear-quadratic regulator algorithm~\cite{friedland2012control} on the lateral deviation in the MSF output, which calculates the amount of steering that will be applied to correct the deviation. We thus mathematically translate such steering into physical position and heading rate changes (detailed in Appendix~\ref{appendix:steer_to_position}), and add them to the original LiDAR locator position and IMU values to get the changed ones due to AD control. The benefit of this method is that it contains real-world sensor noises, which is the key contributor to the take-over vulnerability (\S\ref{sec:security_analysis}). However, it does not model more complicated sensing and vehicle motion factors such as raw LiDAR point cloud changes and tire-road frictions, which thus may have limited synthesizing accuracy.

In the simulation-based evaluation, we directly use an AD simulator to dynamically generate raw sensor inputs to Apollo according to its control decisions in the real time, which has more advanced sensor and vehicle motion modelling. However, a common limitation for AD simulators today~\cite{Dosovitskiy17, lgsvl} is that they do not consider generating sensor data with real-world noises. To address this, we model the LiDAR noises as position errors following a normal distribution with a zero mean for each point of the raw LiDAR point cloud generated from the simulator according to the LiDAR datasheet~\cite{lidardatasheet}.

\textbf{Experimental setup.} In the trace-based evaluation, we run Apollo version 2.5 (the latest version directly compatible with \textit{ba-local})
with the control module enabled on a GPU server, and feed trace \textit{ba-local}. We write a standalone ROS node that feeds the spoofed GPS inputs and also performs the LiDAR locator and IMU input changes described above. For FusionRipper, we use the best attack parameter in \textit{ba-local} with 2-minute minimum attack duration. We do not run the perception module since in Apollo the perception module only outputs detected road obstacles and the system solely relies on the localization module to identify deviations on the road. This is the most popular design modularization for high-level AD systems today~\cite{udacity_av_nd, udacity_av_apollo, coursera_av, apollo, autoware}, which lets the localization module to take charge of all aspects related to vehicle positioning.

In the simulation-based evaluation, we use LGSVL, a production-grade AD simulator that can interface with Apollo version 5.0~\cite{lgsvl}. Since Apollo version 5.0 replaces the ROS runtime with Cyber~\cite{apollo}, we implement the attack logic and noise modeling in a Cyber node instead. Different from the trace-based evaluation, we run the simulation on the \textit{complete} Baidu Apollo AD system with all functional modules enabled, i.e., localization, transform, perception, prediction, planning, routing, and control~\cite{apollo}. 
We simulate two attack scenarios with one attacking to the left of the road and another to the right, where both have concrete safety consequences such as hitting the road barrier or traffic sign.

\textbf{Trace-based evaluation results.} Our results show that FusionRipper achieves 97.0\% and 93.9\% success rates for off-road and wrong-way attacks respectively, which is only slightly lower than those in the MSF algorithm-only analysis (98.0\% and 97.0\%). Such slightly effectiveness drop may be due to run-time randomness when running the end-to-end Apollo system since it uses multi-threading when feeding the sensor inputs to BA-MSF.

\begin{figure}[tbp]
\centering
\includegraphics[width=\columnwidth]{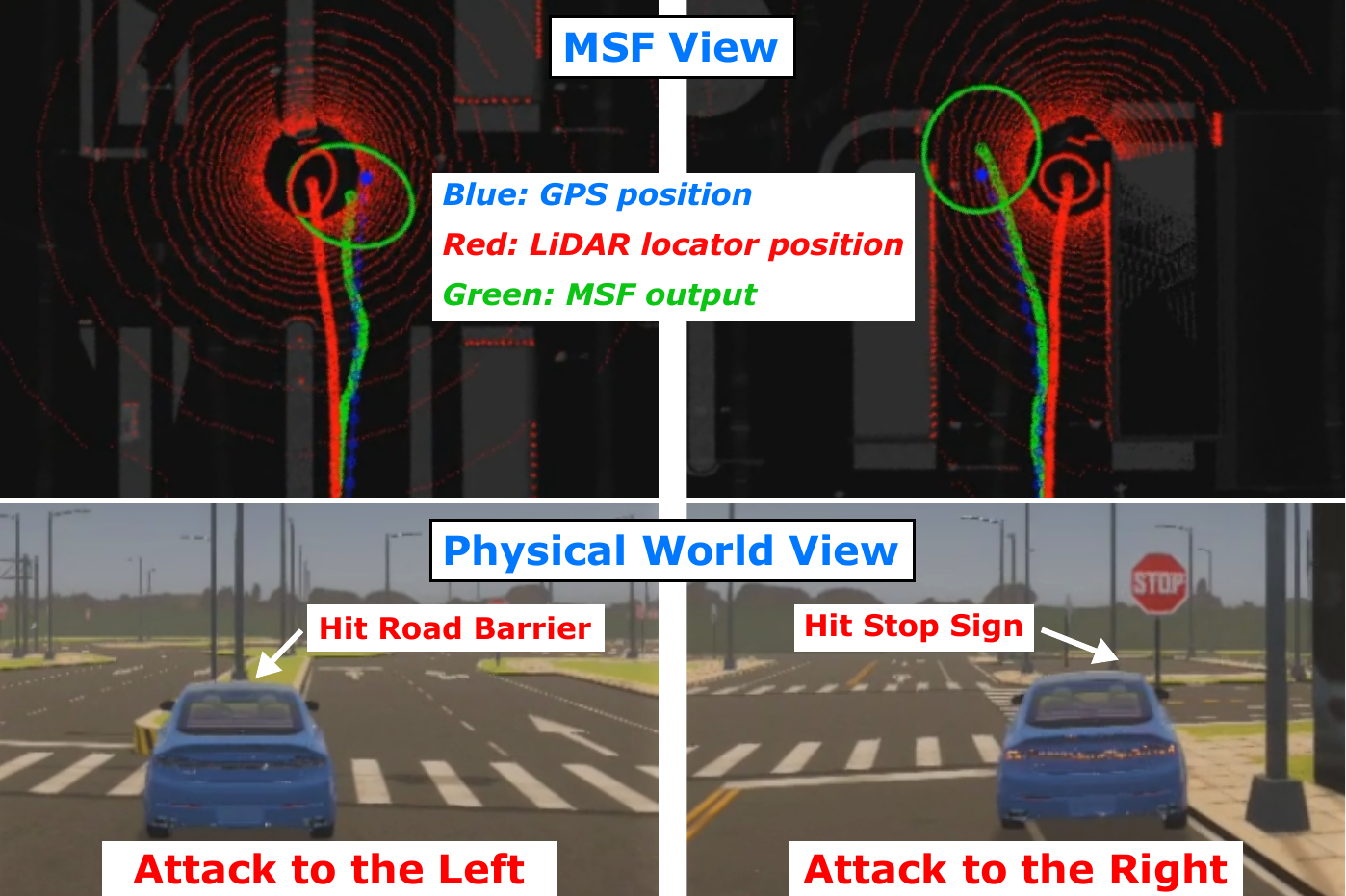}
\vspace{-0.25in}
\ncaption{Snapshots of our end-to-end attack demos~\cite{attack_demo}. MSF View: input sensor positions and MSF outputs; Physical World View: victim AV's physical world position.}
\label{fig:demo}
\vspace{-0.15in}
\end{figure}

\textbf{Simulation-based evaluation results and attack demos.} Our simulation results show that FusionRipper can successfully deviate the victim AV to hit the road barrier or traffic sign even with the complete end-to-end Baidu Apollo AD system operating.
We record attack demo videos for these two simulation scenarios, available at our project website \textbf{\url{https://sites.google.com/view/cav-sec/fusionripper}}. Fig.~\ref{fig:demo} shows a snapshot of the demos. As shown, to correct the MSF output deviation to the right/left of the planned trajectory (i.e., lane center), the AV in the physical world deviates to the left/right and eventually hit the road barrier or the stop sign.

%% file: offline_profile.tex
\vspace{-0.05in}
\nsection{Offline Attack Parameter Profiling}\label{sec:profiling}
\vspace{-0.07in}

Our results so far show that for each trace there always exist an attack parameter combination, \ie, $d$ and $f$, that can achieve high success rates (\S\ref{sec:evaluation}) with high robustness to practical factors (\S\ref{sec:practical_attack}). However, in~\S\ref{sec:effectiveness} we also observe that such high effectiveness is sensitive to the selection of attack parameters. Thus, it is highly desired if there exists an offline method that can efficiently identify highly effective attack parameters before the actual attack. In this section, we thus explore the possibility of designing such a method  to further improve the practicality of FusionRipper.

\vspace{-0.05in}
\nsubsection{Problem Settings and Design} \label{sec:offline_profiling_methodology}
\vspace{-0.05in}

\textbf{Problem Settings.} To find the effective attack parameters offline, we assume that the attacker can perform trials of FusionRipper attacks with different combinations of $d$ and $f$ on AVs of the \textit{same model} as that of the victim AV, \ie, having the same sensor set, AD system, and vehicle model. This is a realistic assumption since any AV models developed for commercial purpose need to be mass produced for the ease of management and reducing the development cost for the self-driving taxi or truck services today~\cite{waymo_production, uber_dallas, lyft_public, baidu-apollo-go}. For example, Waymo's 20,000 self-driving taxis in Phoenix are deployed with the same sensor suite on the same car model~\cite{waymo_20k_av_fleet}, and the same applies to Hyundai's self-driving taxis~\cite{hyundai_socal_av_fleet}.
In this process, the attack trials can be performed \textit{actively}, by requesting the self-driving taxi or truck services that use the targeted AV model, or directly purchasing an AV of the same model.

In such profiling process, it is necessary to prevent causing obvious safety problems both for the attacker's own safety and for remaining stealthy. Thus, in such offline profiling we choose a \textit{safe profiling design}, which still performs the FusionRipper attack but stops the attack right after the physical-world deviation of the AV is over a \textit{safe profiling threshold}. This will thus let the non-spoofed GPS and other positioning sources to drag the MSF output deviations back.

\textbf{Offline profiling algorithm design.} Under the problem settings above, our profiling method is designed following a simple strategy: performing attack trials using different combinations of $d$ and $f$ until we find a combination with a sufficiently high success rate. More specifically, the trials are performed for a number of \textit{profiling rounds}. In each round, the attacker picks one combination of $d$ and $f$ and tries it for multiple times. When picking the combinations, the attacker follows the order from the smallest one to the largest one in the parameter space, since larger ones can more easily make the spoofed inputs outliers and thus directly cause attack failure.

Due to the safety requirement, the attacker follows the safe profiling design above, and considers a $d$ and $f$ combination as successful once it reaches the safe profiling threshold. After each profiling round, the attacker can thus obtain a success rate for a $d$ and $f$ combination. Once the success rate of a combination in a round is over a \textit{minimum profiling success rate}, the profiling terminates and such combination is selected for the actual attack. If the attack parameters space is exhausted, the combination with the highest success rate in profiling is selected. The pseudocode of this method is in Algorithm~\ref{alg:offline_profiling}.

\begin{algorithm}
    \footnotesize
    \caption{Offline Attack Parameter Profiling}
    \label{alg:offline_profiling}
    \textbf{Notations:} \\
    $\text{\attacktrials}(d, f, n, t)$: Profile $n$ attack trials with parameters $d$, $f$, returns the number of trials that have deviations larger than $t$\\
    $N$: Number of attack trials in each profiling round\\
    $S$: Minimum profiling success rate\\
    $T$: Safe profiling threshold \\
    \textbf{Output:} $d$, $f$, \text{cost} \\
    \textbf{Initialize} $d, d_{\text{best}}\gets d_{\text{min}}$; $f, f_{\text{best}}\gets f_{\text{min}}$; $\text{SuccRate}_{\text{best}}, \text{cost}\gets0$
    \begin{algorithmic}[1]
    \ForEach{$f \gets f_{\text{min}}$ to $f_{\text{max}}$}
        \ForEach{$d \gets d_{\text{min}}$ to $d_{\text{max}}$}
            \State \text{SuccCount} $\gets \text{\attacktrials}(d, f, N, T)$
            \State \text{cost} $\gets \text{cost} + N$
            \State \text{SuccRate} $\gets \text{SuccCount}/N$
            \If{$\text{SuccRate} \ge S$}
                \State \textbf{return} $d$, $f$, \text{cost}
            \Else
                \If{$\text{SuccRate} > \text{SuccRate}_\text{best}$}
                    \State $d_{\text{best}} \gets d$, $f_{\text{best}} \gets f$
                    \State $\text{SuccRate}_\text{best} \gets \text{SuccRate}$
                \EndIf
            \EndIf
        \EndFor
    \EndFor
    \State \textbf{return} $d_{\text{best}}$, $f_{\text{best}}$, \text{cost}
    \end{algorithmic}
\end{algorithm}

\vspace{-0.05in}
\nsubsection{Experiments and Evaluation} \label{sec:offline_profiling_expr}
\vspace{-0.05in}

\textbf{Experimental setup.} In this section, we use the 5 KAIST traces used in~\S\ref{sec:effectiveness} since this represents the case with attacking the same AV model (the KAIST traces are collected using the same vehicle on different roads~\cite{jeong2019complex}). We split the 5 traces into two sets, with 4 as the \textit{profiling traces}, \ie, representing the attack trials in the offline profiling, and 1 as the \textit{evaluation trace} for evaluating the selected $d$ and $f$ from profiling, \ie, representing the actual attack on the victim AV. We evaluate all the 5 possible splittings, and then use their average success rate to measure the offline profiling effectiveness. We use the same parameter space as that in~\S\ref{sec:evaluation}.

\textbf{Algorithm parameter choices.} In the profiling algorithm, there are two configurable parameters: \textit{minimum profiling success rate}, and \textit{safe profiling threshold}. Thus, we first perform experiments to understand how to best configure them. In these experiments, for each $d$ and $f$ combination we consider all attack starting points in the profiling traces as its corresponding set of attack trials in the profiling algorithm in order to understand general properties of different parameter values.

We first perform experiments by running the profiling algorithm for different minimum profiling success rates without considering safe profiling design. Our results show that the average success rate of the selected $d$ and $f$ does not change significantly overall. Particularly, it peaks when the minimum profiling success rate is 50\% for both attack goals and drops after that, maybe due to the overfitting to the profiling traces. More details are in Fig.~\ref{fig:min_profile_succ_rate_and_threshold} (a) in the Appendix.

Next, with 50\% as the minimum profiling success rate, we vary the safe profiling threshold, and find that reducing the safe profiling thresholds only slightly changes the average success rate of the selected $d$ and $f$: the success rate differences between profiling threshold 0.3 and 0.9 meters are less than 4\% for both attack goals. In particular, using 0.45 meters as the safe profiling threshold has the overall highest average success rate for both attack goals, which are 90.3\% and 84.4\% respectively. Details are in Fig.~\ref{fig:min_profile_succ_rate_and_threshold} (b) in the Appendix. Such 0.45 meters deviation does not cause the AV to drive off road on both local roads and highway (Table~\ref{tbl:goal_deviations}). On local roads, it will only cause very slightly lane straddling, and on the highway, it is far from even touching the left or right lane line (both visualized in Fig.~\ref{fig:visualize_safe_threshold} in Appendix). Thus, the attacker can choose to perform such safe profiling on the highway, or on the local roads with light traffic.

\textbf{Evaluation results.} With the algorithm parameter values decided, we then evaluate the algorithm effectiveness and the profiling cost with limited number of attack trials for each combination of $d$ and $f$ in the profiling round. We define profiling cost as the total number of attack trials spent in the profiling algorithm, since in our problem setting each trial corresponds to a self-driving trip the attacker needs to take, \eg, from a targeted self-driving taxi service. For each attack trial, we limit its maximum duration to 90 seconds, which generally covers over 95\% of the successful cases according to our earlier evaluation on attack success time (\S\ref{sec:effectiveness}).

\begin{figure}[tbp]
\centering
\includegraphics[width=.7\columnwidth]{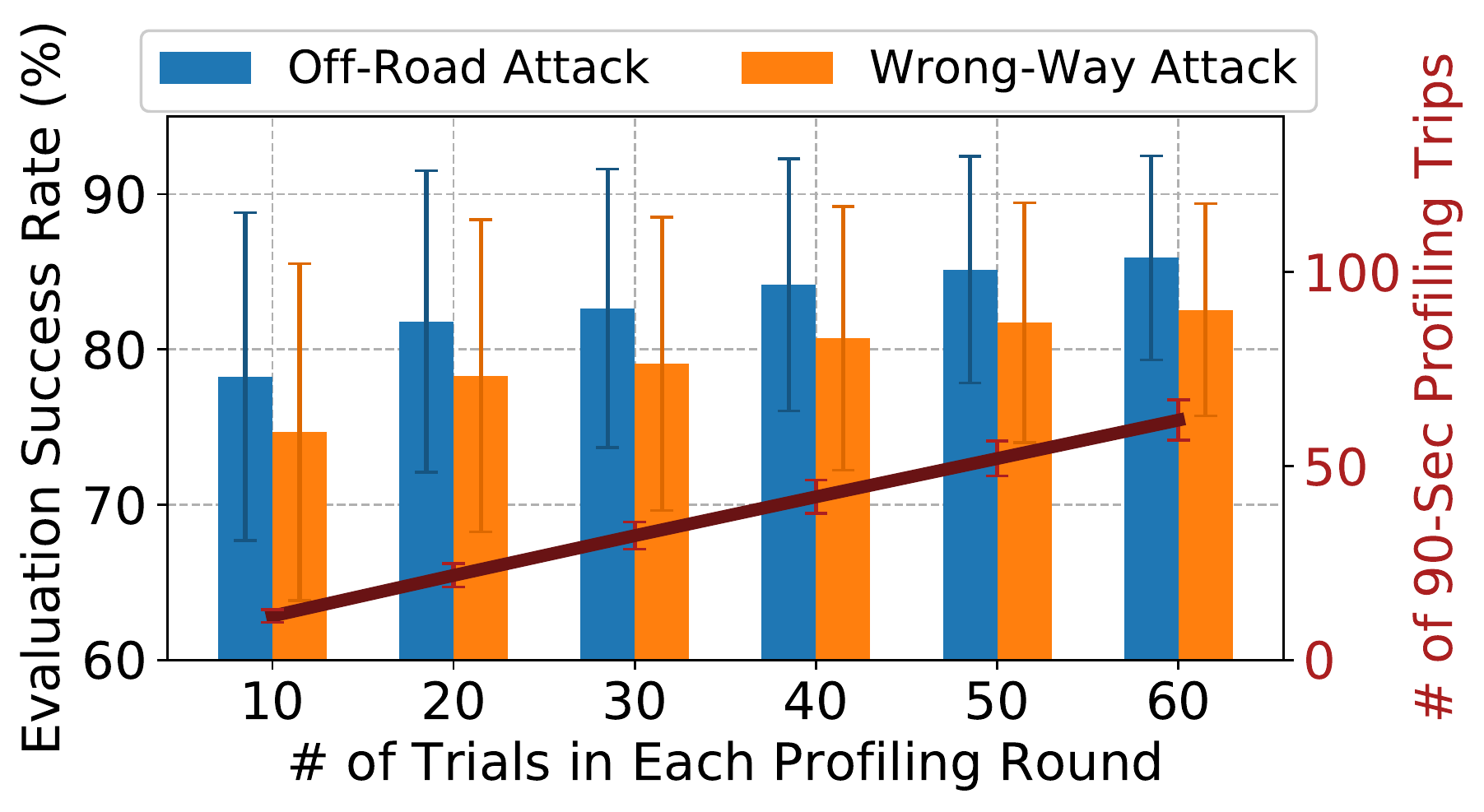}
\vspace{-0.15in}
\ncaption{Average profiling effectiveness (bar graph) and costs (line graph) under different numbers of attack trials in each profiling round. Each profiling is repeated for 100 times.}
\label{fig:profiling_cost}
\vspace{-0.15in}
\end{figure}

Fig.~\ref{fig:profiling_cost} shows the average success rates of the $d$ and $f$ output by the profiling algorithm and the average numbers of 90-sec profiling trips under different numbers of attack trials in each profiling round. In each profiling round, we randomly sample the corresponding number of attack trials from all attack starting points in the profiling traces. As shown, the average success rate increases as the attacker spends more trials in each profiling round since with more trials, the profiled success rate of a $d$ and $f$ combination in a profiling round is statistically closer to the ground truth. Particularly, when the number of trials in each profiling round is 40, our profiling algorithm can find a $d$ and $f$ combination with over 80\% average success rate for both off-road and wrong-way attacks (84.2\% and 80.7\% respectively). In this case, \textit{the profiling cost is only 42 1.5-minute trips on average, which in total is only slightly over 1 hour}. Since the attackers can actively perform such trials, \eg, by requesting self-driving taxi services themselves, finishing this should take at most half a day.

%% file: discussion.tex
\vspace{-0.05in}
\nsection{Limitation and Defense Discussions}\label{sec:discussion}
\vspace{0.07in}

\nsubsection{Limitations of Our Study} \label{sec:limits}
\vspace{-0.05in}

\textbf{Study representativeness.} As the first work to study the security of MSF-based AD localization, we choose to focus on the most representative design, KF-based MSF, and the most representative implementation we can find, BA-MSF (representativeness discussed in~\S\ref{sec:msf}). However, it is still unclear whether other less common MSF designs (\eg, particle filter based~\cite{zhang2019localization}) and outlier detection designs (\eg, expectation-maximization based~\cite{ting2007kalman}) can be more secure, which can be potential future work directions.

\textbf{Attack generality.} Although our results have shown the generality of FusionRipper by showing high success rates on 3 different KF-based MSFs (\S\ref{sec:generality}), only one (BA-MSF) of them is production-grade implementation for AD systems. Ideally it is better to evaluate on other production-grade ones, but very unfortunately BA-MSF is the only one that is publicly available so far and it is unlikely for other AV companies to publicly release their implementations in the near future. Thus, due to the lack of information, it is unclear whether other leading AV companies, e.g., Waymo and GM, are vulnerable to our attack. Nevertheless, since BA-MSF is representative both at the design and implementation levels (\S\ref{sec:msf}) and our attack is general to KF-based MSF by design (\S\ref{sec:cause_analysis}), if other AV companies also adopt such a representative design, at least at design level they are also susceptible to the discovered take-over vulnerability. Thus, as the first study, we believe our current discovery and evaluation results can already most generally benefit the understanding of the security property of MSF-based AD localization today.

\textbf{Attack practicality.} We evaluate FusionRipper on real-world traces and under various practical factors such as spoofing inaccuracies and AD control taking effect (\S\ref{sec:practical_attack}). To further improve the attack practicality, we design an offline attack parameter profiling method that can achieve 84.2\% and 80.7\% success rates for off-road and wrong-way attacks, with the profiling cost of at most half a day. Nevertheless, due to the cost and legal regulation for GPS spoofing, we did not conduct attack experiments on real-world AVs, which thus can be a valuable future work. Note that GPS spoofing has been proven practical on various end systems~\cite{popperccs11, utaustinspoofer, zeng2018all, narain2018security, franceschi2012drone, kerns2014unmanned, spoof_tesla, spoofyacht}, including cars such as Tesla cars~\cite{spoof_tesla} (\S\ref{sec:spoofing}). Moreover, in this work, we model GPS spoofing based on attack capabilities shown in prior work~\cite{zeng2018all, narain2018security, spoofyacht} to minimize any unrealistic assumptions. 

As mentioned in \S\ref{sec:threat_model_details}, we assume the attacker owns an AV and can leverage AD perception algorithms to track the physical position of the victim. Although accurate position-tracking of surrounding obstacles is a basic task for AVs, we did not conduct physical-world experiments to confirm this, which is thus left as a valuable future work.

\vspace{-0.05in}
\nsubsection{Defense Discussions} \label{sec:defense}
\vspace{-0.05in}

In this section, we discuss the potential defense directions against FusionRipper.

\textbf{Defend against GPS spoofing.} Our attack depends on GPS spoofing, so one direct defense direction is to leverage existing GPS spoofing detection or prevention techniques. Unfortunately, neither GPS spoofing detection nor prevention are fully-solve problems today. On the detection side, numerous techniques have been proposed
leveraging signal power monitoring~\cite{akos2012s, psiaki2016gnss, ranganathan2016spree}, multi-antenna based signal arrival angle detection~\cite{magiera2015detection, psiaki2016gnss}, or crowdsourcing based cross-validation~\cite{jansen2018crowd}.
However, they either can be circumvented by more advanced spoofers~\cite{kerns2014unmanned, psiaki2016gnss} or are only applicable to limited domains such as airborne GPS receivers~\cite{jansen2018crowd}.
On the prevention side, cryptographic authentication based civilian GPS infrastructure can fundamentally prevent direct fabrications of GPS signals~\cite{psiaki2016gnss}. However, it requires significant modifications to the existing satellite infrastructure and GPS receivers, and is still vulnerable to replay attacks~\cite{papadimitratos2008gnss}. Thus, one interesting future work direction is to more concretely understand how effective the latest GPS spoofing defense techniques can be against the current or adapted versions of FusionRipper.

\textbf{Improve confidence of MSF state and LiDAR locator.} Another fundamental defense direction is to improve the positioning confidence of MSF state and LiDAR locator, the two most important factors to the take-over vulnerability in real-world trace (\S\ref{sec:security_analysis}). Fundamentally, such lacks of confidence in practice result from algorithm inaccuracies and sensor noises (\S\ref{sec:security_analysis}), and as shown in our analysis, even for the high-end sensors and production-grade LiDAR locator used in AVs today, these inaccuracies and noises are unfortunately large and frequent enough for FusionRipper to exploit. To improve on this, substantial technology breakthrough in sensing and LiDAR-based localization needs to take place. Unfortunately, it is unclear when such breakthrough can take place.

\textbf{Leverage independent positioning sources (\eg, camera-based lane detection) as fail-safe features for high-level AD localization.}
Since fundamental defense directions above are not immediately deployable, it is highly desired to discuss the possibility of short-term mitigation solutions. One promising direction is to leverage independent positioning sources to cross-check the localization results and thus serve as \textit{fail-safe} features for AD localization. For example, since both off-road and wrong-way attacks will cause the victim AV to deviate from the current lane, they should be detectable by \textit{camera-based lane detection}~\cite{hillel2014recent}, a mature technology available in many vehicle models today~\cite{lane_keeping_on_cars}. However, we find that in the high-level AD system design today, such a technology has not been generally considered for fail-safe purposes. For example, the latest release of Baidu Apollo (version 5.5) uses it only for camera calibration~\cite{apollo_perception}, while Autoware does not use it at all~\cite{autoware_perception}.
This might be because the lane detection output is local positioning within the current lane boundaries, and thus cannot be directly used for comparison against global positioning from MSF.
However, the vulnerability discovered in this paper strongly motivates the need for considering adding such kind of fail-safe features in future AD localization, at least for \textit{anomaly detection}.
Note that more investigations are needed to understand how effective and robust such kind of fail-safe features can be in the defense. For example, when camera-based lane detection is applied for anomaly detection, the precision/recall rates need to be further explored since it needs to carefully consider (1) AVs legitimately deviating from current lane due to routing requirements, and (2) lane line scratches or incompleteness. Moreover, camera-based lane detection itself is vulnerable to physical-world attacks~\cite{tencent2019, sato2020security}.

Note that even if such fail-safe features can perform perfect attack detection, our attack still causes denial-of-service of the victim's global localization function, which can render the victim in unsafe scenarios, e.g., stopping in the middle of highway lanes, since the victim can neither correctly reach the destination nor safely locate the road shoulder to pull over. Thus, a more useful defense direction is to \textit{correct} the attacked localization results. However, so far the global positioning accuracy of cameras is unsatisfying for high-level AD localization, especially along the longitudinal direction (forward/backward) since only the stop lines can be used as features~\cite{chong2013synthetic, lee2015gps}. This is why LiDAR locator is used more predominantly in high-level AD localization (\S\ref{sec:msf}). Moreover, such correction is yet another multi-sensor fusion problem and thus is still fundamentally vulnerable to the take-over vulnerability discovered in this paper (\S\ref{sec:security_analysis}). Thus, how to leverage other independent positioning sources to effectively perform such correction under our attack is still an open research challenge, which can be a valuable future work direction.

%% file: related_work.tex
\vspace{-0.05in}
\nsection{Related Work} \label{sec:related_work}
\vspace{-0.07in}

\textbf{GPS spoofing on navigation systems.}
Recently, Zeng~\etal~\cite{zeng2018all} find that GPS spoofing can be used to stealthily deviate a victim car to an attacker-controlled destination. Later Narain~\etal~\cite{narain2018security} further find that such attack also exists for a GPS/INS (Inertial Navigation System) navigation system.
Compared to our work on MSF-based localization, these prior works target single-source localization systems without fusion from other position sources, such as a LiDAR locator.

\textbf{Theoretical work on KF security.}
Existing theoretical works~\cite{su2016stealthy,liu2012cyber,mo2010false1,mo2010false2} from the control systems domain have studied the security of KF under sensor spoofing. Compared to our work, they only study single-source KFs without any sensor fusion. Also, they focus on the theoretical aspect of the KF and assume the attacker has full access to the KF internals, \eg, KF state and uncertainties. In comparison, our work does not make such assumptions and hence is much more realistic.

\textbf{AV-related attacks and defenses.}
Various previous works studied security problems on traditional vehicle systems~\cite{checkoway2011comprehensive, baker2019losing, garcia2016lock}, but not AD systems. Closer to this work, prior works discovered various sensor attack vectors on sensors related to AD systems, such as camera, LiDAR, IMU, radar, and ultrasonic sensors~\cite{petit2015remote, yan2016can, cao2019adversarial, tu2018injected, son2015rocking, trippel2017walnut}. However, none of them considers how to leverage these attack vectors to attack AD localization.
On the defense side, recently Choi~\etal~\cite{choi2018detecting} and Quinonez~\etal~\cite{savior} propose to use control or physical invariants to detect sensor attacks to small robotics vehicles such as drones and ground rovers. However, it is unclear how these methods can be effectively applied to AD systems, since AVs operate in highly complex and dynamic road conditions where the baseline/normal behaviors can be much harder to accurately model or predict.

%% file: conclusion.tex
\vspace{-0.05in}
\nsection{Conclusion}\label{sec:conclusion}
\vspace{-0.07in}

In this paper, we perform the first security study on MSF-based localization in high-level AV settings under GPS spoofing. We discover a take-over vulnerability that can fundamentally defeat the MSF design principle, and design FusionRipper, a novel and general attack that opportunistically captures and exploits it. Our evaluation on real-world traces shows that FusionRipper can achieve over 97\% and 91.3\% success rates in all traces for off-road and wrong-way attacks. Such high effectiveness is also found highly robust to various practical factors. We also design an offline method that can identify effective attack parameters within at most half a day. We also discuss both long-term and short-term defenses directions, and identify that a promising mitigation is to use camera-based lane detection as a fail-safe feature, which has not been generally considered for such purpose today. As the first study on AD localization security, we hope that our findings and insights can bring immediate attention and inspire the development of effective defenses considering the critical role of localization for safe and correct AV driving.

%% file: acknowledgement.tex
\vspace{-0.1in}
\section*{Acknowledgments}
\vspace{-0.1in}

We would like to thank Takami Sato, Ningfei Wang, Ziwen Wan, Shinan Liu, Alex Veidenbaum, Gene Tsudik, Marco Levorato, Ardalan Amiri Sani, Joshua Garcia, Yu Stephanie Sun,
the anonymous reviewers, and our shepherd, Yongdae Kim, for providing valuable feedback on our work. This research was supported in part by the National Science Foundation under grants CNS-1850533 and CNS-1929771.

%% file: appendix.tex
\begin{table}[tbp]
\footnotesize
\centering
\ncaption{Notations in KF and contributing factor derivation.}
\label{tbl:notations}
\vspace{-0.1in}
\begin{tabular}{|c|l|}
\hline
\textbf{Notation} & \textbf{Description} \\ \hline
$\mathbf x$ & KF state, \eg, PVA \\ \hline
$\mathbf P$ & KF state co-variance, \ie, state uncertainty \\ \hline
$\mathbf K$ & Kalman gain of the measurement \\ \hline
$\mathbf F$ & \begin{tabular}[c]{@{}l@{}}State transition model;\\
it describes the kinematics functions used in KF prediction\end{tabular} \\ \hline
$\mathbf H$ & \begin{tabular}[c]{@{}l@{}}Observation model; it is an identity matrix\\ if the measurement and state have the same scale\end{tabular} \\ \hline
$\mathbf Q$ & \begin{tabular}[c]{@{}l@{}}Process noise co-variance;\\ usually a fixed pre-tuned matrix\end{tabular} \\ \hline
$\mathbf z$ & Sensor measurement \\ \hline
$\mathbf R$ & Measurement variance, \ie, measurement uncertainty \\ \hline
$\mathbf r$ & Victim's physical position \\ \hline
$\delta$ & Spoofing distance to victim's physical position \\ \hline
$\Delta$ & LiDAR position distance to the original MSF position \\ \hline
$dev$ & The deviation after each KF operation under spoofing \\ \hline
\end{tabular}
\vspace{-0.1in}
\end{table}

\section{Calculation of Required Deviations in Attack Goals and Distances to Lane Line}\label{appendix:goal_deviations}

\begin{figure}[tbh]
    \begin{minipage}[tbh]{0.47\linewidth}
        \centering
        \includegraphics[width=.95\columnwidth]{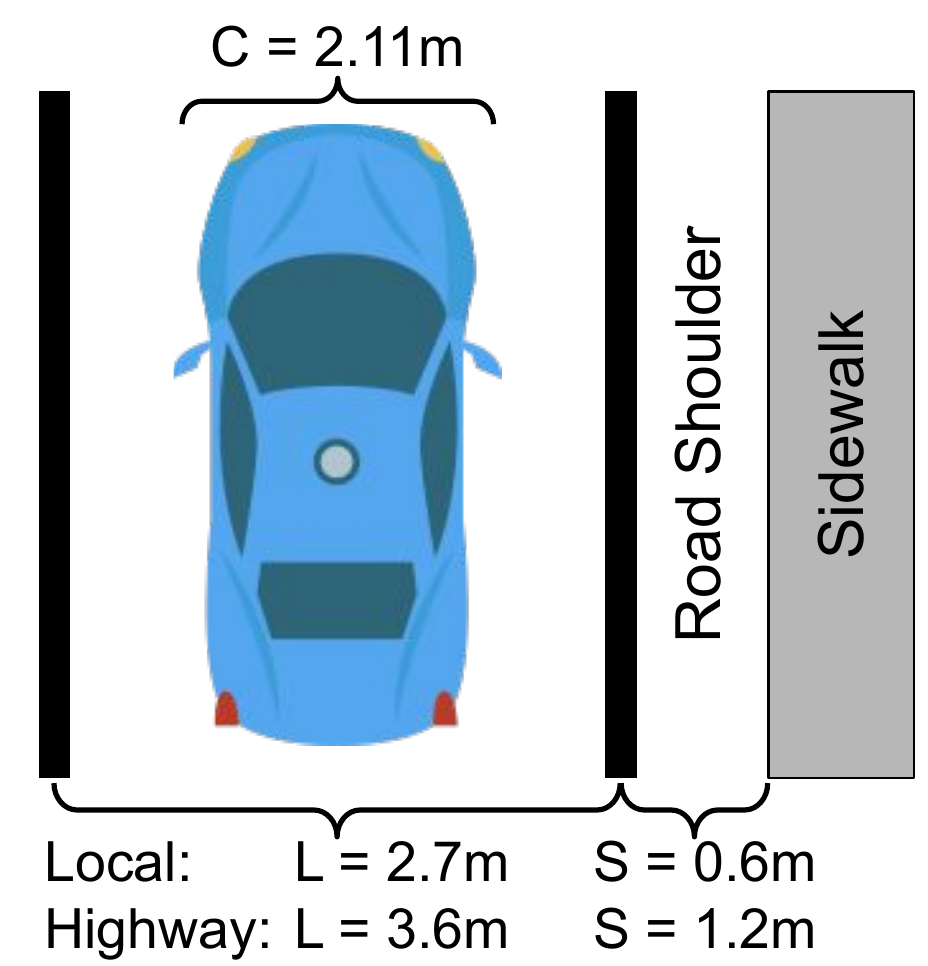}
        \caption{Common AV, traffic lane, and road shoulder widths used in this paper.}
        \label{fig:std_measurements}
    \end{minipage}
    \hspace{0.01\linewidth}
    \begin{minipage}[tbh]{0.50\linewidth }
        \centering
        \includegraphics[width=.95\columnwidth]{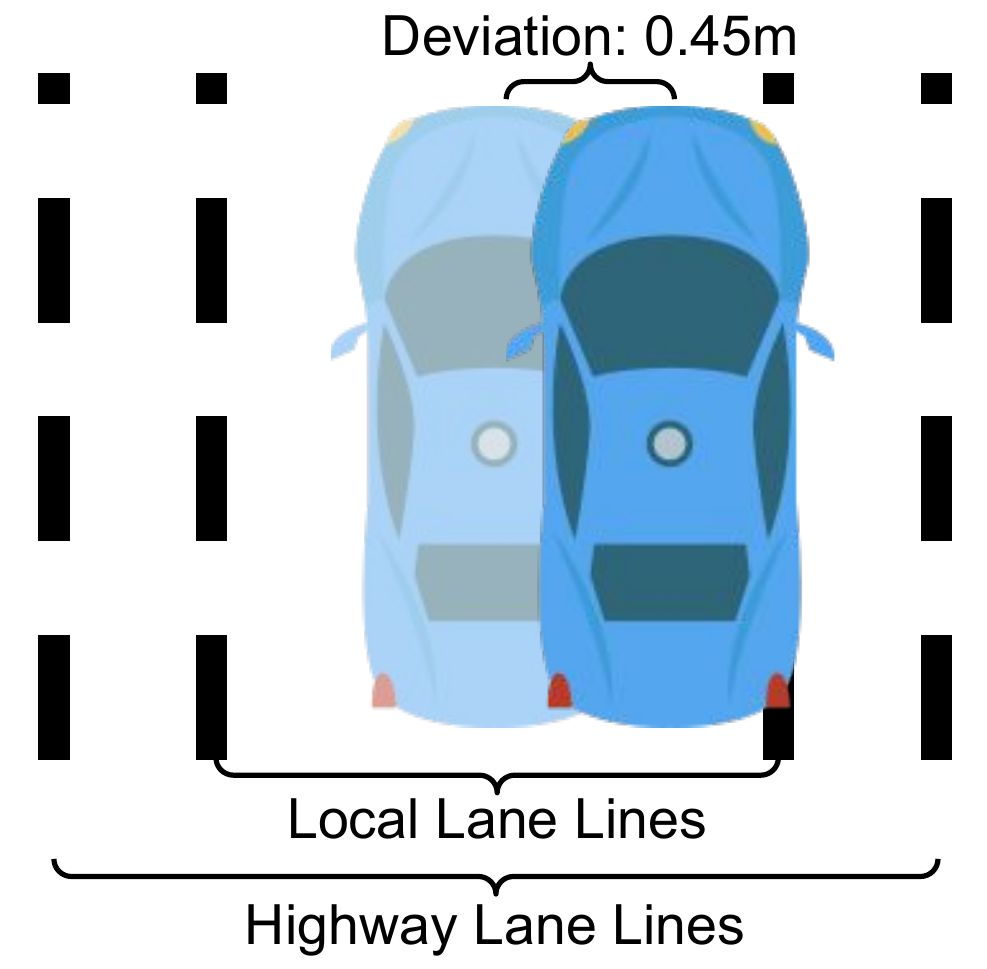}
        \caption{Visualization of the lateral deviation 0.45 meters on local and highway roads.}
        \label{fig:visualize_safe_threshold}
    \end{minipage}
\vspace{-0.15in}
\end{figure}

The required deviations under off-road and wrong-way attacks are calculated based on common widths of the AV, lane, and the road shoulder. These values differ in local and highway settings. Fig.~\ref{fig:std_measurements} shows the width measurements we used in the calculation. For the AV width, we use the width (including mirrors) of the Baidu Apollo's reference car, Lincoln MKZ~\cite{mkz_spec}. For the lane widths and shoulder widths, we refer to the design guidelines~\cite{road_shoulder_width} published by the US Department of Transportation Federal Highway Administration. For off-road attack, we use the deviation when the AV goes \textit{beyond} the road shoulder from the center of the lane as the required deviation, which is calculated using $\frac{L-C}{2}+S = 0.895m$~(local) and $1.945m$~(highway), where $L$ is the lane width, $C$ is the car width, and $S$ is the road shoulder width. For wrong-way attack, we define the required deviation as the AV completely invades the neighbor lane, and it is calculated with $\frac{L}{2}+\frac{C}{2} = 2.405m$~(local) and $2.855m$~(highway). We calculate the deviation of \textit{touching the lane line} using $\frac{L-C}{2}$, which is $0.295m$ on local roads and $0.745m$ on the highway.

\section{Convert Steering to Lateral Position and Heading Rate Changes}\label{appendix:steer_to_position}

\begin{figure}[tbp]
\centering
\includegraphics[width=.65\columnwidth]{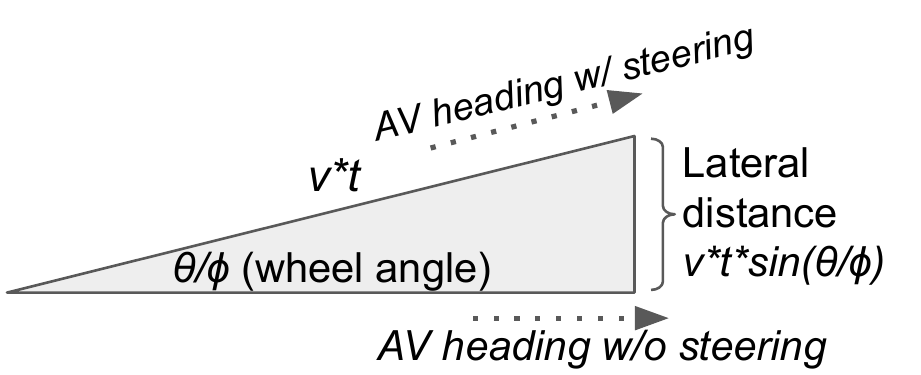}
\vspace{-0.05in}
\ncaption{Conversion from the steering wheel angle to lateral position change.}
\label{fig:steer_to_position}
\vspace{-0.15in}
\end{figure}

Fig.~\ref{fig:steer_to_position} shows the mathematical conversion from the steering angle to physical world lateral position change. The position change can be calculated as $\delta_{\text{pos}} = vt\sin(\frac{\theta}{\phi})$, where $v$ is the velocity, $t$ is the cycle time of the controller, $\theta$ is the steering angle, and $\phi$ is the steering ratio, which is a constant describing the ratio of the turning angle of the steering wheel to that of the vehicle wheel.
The steering angle can be directly converted to heading rate change using $\delta_{\omega} = \theta/\phi t$, where $\delta_{\omega}$ is the yaw (\ie, heading) rate change.

\begin{figure}[tbp]
\centering
\includegraphics[width=.95\columnwidth]{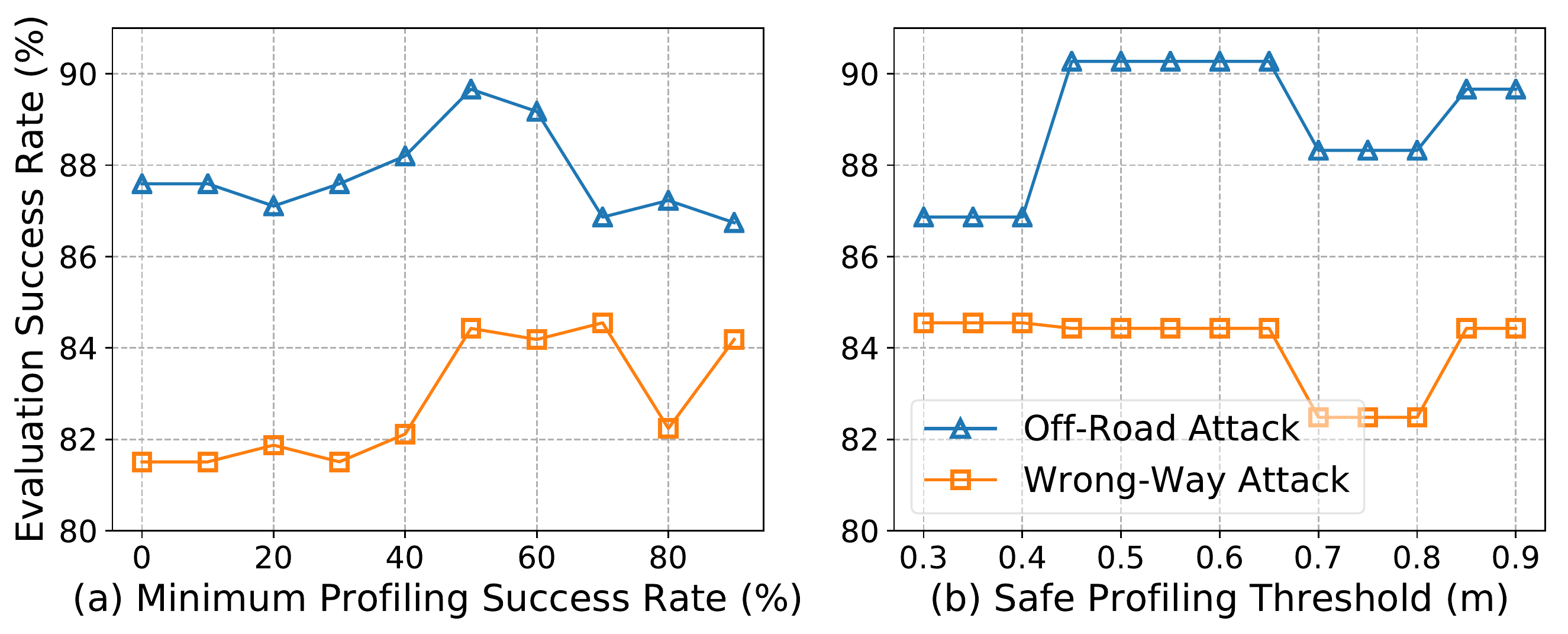}
\ncaption{Profiling results when using different (a) minimum profiling success rates, and (b) safe profiling thresholds.}
\label{fig:min_profile_succ_rate_and_threshold}
\end{figure}